\documentclass[lettersize,journal]{IEEEtran}
\usepackage{amsmath,amsfonts}
\usepackage{algorithm}
\usepackage{array}
\usepackage{textcomp}
\usepackage{stfloats}
\usepackage{url}
\usepackage{verbatim}
\usepackage{graphicx}
\usepackage{cite}
\usepackage{graphicx}%
\usepackage{multirow}%
\usepackage{amsmath,amssymb,amsfonts}%
\usepackage{amsthm}%
\usepackage{mathrsfs}%
\usepackage{xcolor}%
\usepackage{textcomp}%
\usepackage{manyfoot}%
\usepackage{booktabs}%
\usepackage{algpseudocode}%
\usepackage{listings}%
\usepackage{subcaption}
\usepackage[hidelinks]{hyperref}
\usepackage[capitalize]{cleveref}
\usepackage{amsmath}
\usepackage{tcolorbox}
\usepackage{tabularx}
\usepackage[nolist]{acronym}
\usepackage{lineno}
\usepackage{caption}
\usepackage{subcaption}
\newcommand{\norm}[1]{\left\lVert#1\right\rVert}
\begin{document}

\title{Empowering Prosumers: Incentive Design for Local Electricity Markets Under Generalized Uncertainty and Grid Constraints}

\author{
    Pål~Forr~Austnes,~Matthieu~Jacobs,~Lu~Wang,~and~Mario~Paolone,~\IEEEmembership{Fellow,~IEEE}\\
    Distributed Electrical Systems Laboratory (DESL), École~Polytechnique~Fédérale~de~Lausanne~(EPFL), 1015 Lausanne, ~Switzerland\\
    Email: \{pal.austnes, matthieu.jacobs, lu.wang, mario.paolone\}@epfl.ch
}




\maketitle

\begin{acronym}
    \acro{DR}{Demand Response}
    \acro{SM}{Smart Meter}
    \acro{ADN}{Active Distribution Network}
    \acro{D-PMU}{Distribution-Phasor Measurement Unit}
    \acro{LEM}{Local Electricity Market}
    \acro{CC-OPF}{Chance-constrained Optimal Power Flow}
    \acro{gPC}{general Polynomial Chaos}
    \acro{HEMS}{Home Energy Management System}
    \acro{DG}{Distributed Generator}
    \acro{SOC}{State of Charge}
    \acro{DER}{Distributed Energy Resource}
    \acro{ESS}{Energy Storage System}
    \acro{TSO}{Transmission System Operator}
    \acro{ISO}{Independent System Operator}
    \acro{DSO}{Distribution System Operator}
    \acro{LMP}{Locational Marginal Price}
    \acro{PLMP}{Probabilistic Locational Marginal Price}
    \acro{PV}{Photo-Voltaics}
    \acro{DP}{Dynamic Programming}
    \acro{SOE}{state-of-energy}
    \acro{CEM}{Central Electricity Market}
    \acro{GCP}{grid connection point}
    \acro{EV}{Vlectric Vehicle}
    \acro{LV}{Low Voltage}
    \acro{MV}{Medium Voltage}
    \acro{GHI}{Global Horizontal Irradiance}
    \acro{MPC}{model predictive control}
    \acro{MO}{Market Operator}
    \acro{UL}{upper layer}
    \acro{LL}{lower layer}
\end{acronym}

\begin{abstract}
Since the 1990s, widespread introduction of central (wholesale) electricity markets has been seen across multiple continents, driven by the search for efficient operation of the power grid through competition. Fueled by the need of reducing green house gas emissions, the last years have seen an exponential increase in electricity generation from renewable sources, in particular from wind turbines and solar power plants. This increase has made significant impacts both on central electricity markets and distribution-level grids as renewable power generation is often connected to the latter. These stochastic renewable technologies have both advantages and disadvantages. On one hand they offer very low marginal cost and carbon emissions, while on the other hand, their output is uncertain, requiring flexible backup power with high marginal cost. Flexibility from end-prosumers or smaller market participants is therefore seen as a key enabler of large-scale integration of renewables. However, current central electricity markets do not directly include uncertainty into the market clearing and do not account for physical constraints of distribution grids. In this paper we propose a local electricity market framework based on probabilistic locational marginal pricing, effectively accounting for uncertainties in production, consumption and grid variables. The model includes a representation of the grid using the lindistflow equations and accounts for the propagation of uncertainty using \ac{gPC}. The lindistflow equations combined with \ac{gPC} allows to derive a convex, second-order cone model, ensuring global optimality. A two-stage model is proposed; in the day-ahead stage, probability distributions of prices are calculated for every timestep, where the expected values represents the day-ahead (spot) prices. In the real-time stage, uncertainties are realized (measured) and a trivial calculation reveals the real-time price. Through four instructive case-studies we highlight the effectiveness of the method to incentivize end-prosumers' participation in the market, while ensuring that their behavior does not have an adverse impact on the operation of the grid. The proposed methodology significantly reduces the needs for performing real-time calculations, ensuring its practicality and efficiency in real-world applications.
\end{abstract}


\section*{Nomenclature}
\begin{tcolorbox}[colback=white, colframe=black]
\begin{tabularx}{\textwidth}{l X}
$\mathcal{N}$        & Set of buses \\
$\mathcal{L}$        & Set of branches \\
$\mathcal{K}$        & Set of Polynomial Chaos (PC) coefficients \\
$A$                  & Reduced branch-bus incidence matrix \\
$r, x$               & Branch resistance and reactance \\
$P_k^n, Q_k^n$       & k-th PC-coefficient of the n-th branch active and reactive power flow \\
$p_k^n, q_k^n, v_k^n$       & k-th PC-coefficient of the n-th bus active and reactive power injections and voltage \\
$\lambda_k^n$       & k-th PC-coefficient of the n-th bus dual variable of the active power balance \\
$\mu_k^n$       & k-th PC-coefficient of the n-th bus dual variable of the reactive power balance \\
$\boldsymbol{\xi}$       & Vector of stocastich germ \\
$K$       & Number of PC coefficients \\
$d_\xi, p_\xi$       & Number of elements in stochastic germ and polynomial degree of the expansion \\

\end{tabularx}
\end{tcolorbox}

\section{Introduction}\label{sec1}
\acresetall
\subsection{Central electricity markets and transmission networks}
\acp{CEM}\footnote{In this article, \acp{CEM} refers to country/region-wide wholesale electricity markets, usually operated at the transmission-level of the power grid. Depending on the region, the terminology differs, for example National Electricity Market (Australia) or Wholesale Market (Europe/North America). The term \textit{central} is not to be confused with \textit{centralized}, which refers to the market design.} have expanded greatly since the early beginnings in the 1990s, replacing vertically integrated grids. Their expansion was driven by the well established idea that increasing competitiveness between producers and consumers leads to a more efficient operation, both in terms of costs and reliability. \acp{CEM} vary in how they account for physical constraints of the power grid. Zonal markets do not account for the physics of the power grid within the zone, requiring redispatch if market results are grid-infeasible. Nodal markets usually account for grid physics through a simplified version of the power flow equations. In \acp{CEM}, day-ahead wholesale markets are the most known and, usually, largest ones, aiming to schedule electricity production and, in case of demand response, consumption for the next day. Furthermore, since the power grid requires constant balance between production and consumption, additional operational measures are needed to account for forecast errors, plant failure or transmission grid assets failure. In this respect, the umbrella term \textit{balancing services} covers the provision of these measures, which are often traded in balancing markets.

Direct access of an end-prosumer to \acp{CEM} varies widely across different regions. Notable examples are the European markets, where aggregators trade in the markets and end-prosumers can choose their tariff, for example fixed or time-varying ones. In some other regions, end-prosumers trading a certain volume can participate, while smaller end-prosumers are captive and pay a tariff given by their local supplier.

Since the introduction of electricity markets, the sources of production have shifted significantly. Historically, the rationale for day-ahead markets was to forecast load, perform traditional power plants unit-commitment and give the grid operator time to assess feasibility of market results ex ante. Today, stochastic renewables are increasingly dominating power grids requiring increased balancing to maintain supply/demand equilibrium. In addition, the production from stochastic renewables, such as wind and solar power can only be estimated using weather forecasts along with a detailed knowledge of location and characteristics of distributed generators. The root-mean-squared errors of day-ahead forecasts for wind and solar power are on the order of 5-20\% (\cite{tsai_review_2023, theocharides_impact_2021, wang_cost_2022}) and thus, when their bulk production share becomes significant, the impact of forecasting errors becomes significant too. Therefore the increasing use of probabilistic methods to quantify this uncertainty has been largely developed and employed. Methods producing quantile forecasts or probability density forecasts are now customary when operating such assets. Currently, these probabilistic forecasts are mostly used to perform proper risk management of portfolios, adequacy-studies by grid operators and planning purposes. Their direct inclusion in electricity markets is limited, with a notable exception of the "P90"-requirement in Denmark\footnote{The P90 requirement is a pre-qualification rule for stochastic flexible assets that wish to participate in ancillary-services markets in Denmark. The resource must have the offered capacity available at least in 90\% of the time.} \cite{gade_leveraging_2024}. Some authors have proposed adapting electricity markets to accept probabilistic bids to better account for the inherent uncertainty of the generation and consumption \cite{papakonstantinou_information_2016}.

Balancing the stochasticity of renewables requires dispatchable power generators, such as hydro power, combined cycle gas turbines, batteries, demand response and, to a lesser extent, coal and nuclear power. As the share of renewables grows, the hours where dispatchable power is needed reduces, compressing the economic viability of such plants. At the same time, certain weather-events result in very low production from renewable resources and it might become difficult to serve demand in those critical hours. For example, the \textit{dunkelflaute} is well-known in Europe characterized by periods of 1-14 days with very little sunlight and wind, calling into question an all-renewables supplied power grid. Solving this problem only by increasing the capacity of dispatchable generation is challenging because they are only needed for a limited number of hours per year, effectively making them extremely expensive to operate (e.g. \cite{schwarz_security_2022}).

\subsection{Distribution networks and local electricity markets}
An alternative to the over-dimensioning of reserve capacity is to incite loads to more closely follow renewables generation, through developing \ac{DR} schemes. As known, DR allows for adjusting the consumption of participating loads to better match various grid conditions, such as frequency regulation, redispatch etc. For example, electric boilers can adjust their consumption and temporarily use stored heat to maintain their service. Other types of electricity consumption control, such as electric vehicle charging, can be performed when supply is ample, reducing the need for expensive dispatchable generation. DR of end-prosumers happens in distribution grids, which have historically been passive, i.e., built to serve the grid load under any situation. Driven by similar economic arguments as for \acp{CEM}, \acp{LEM} have been proposed to improve the  operation of distribution grids.\footnote{\acp{LEM} and improved operation of distribution grids are supported by new tools, such as \acp{SM}, \acp{D-PMU} and \acp{ADN}, rendering the grid observable and controllable \cite{noauthor_development_2011}.} Several LEM schemes have been proposed in the literature, such as peer-to-peer and pooled markets \cite{faia_local_2024,bjarghov_developments_2021}. Also several pilot projects have been established \cite{ableitner_quartierstrom-implementation_2019,chondrogiannis_local_2022}. Although the research is promising, complexity of the chosen market clearing mechanism and lack of prosumer participation has rendered these projects difficult to expand. \acp{LEM} naturally have many more end-users and a more complex grid infrastructure than \acp{CEM}, emphasizing the need of scalable and transparent market models.

% To reduce the need for grid expansion, and thus costs, the operation of electricity markets should align more closely with the physical reality of the grid. Accurate price signals can be a proxy for demand response, inciting consumers to shift their consumption to periods of low prices.

Distribution grids are fundamentally different from transmission grids and, therefore, LEMs require different methods to account for grid constraints such as lines/transformers congestions and voltage bounds. In addition, the uncertainty of demand and DERs require methods to optimize under uncertainty. In this regard, the two main paradigms for optimization under uncertainty in power systems relies on robust and chance-constrained approaches. In a robust approach, one aims to minimize the worst case cost, satisfying constraints robustly, i.e. using uncertainty sets along with the base grid constraints. This framework is useful when guarantees of constraint-satisfaction are strictly necessary and the operator must guarantee under any modeled scenario that the overall system constraints and supply security are satisfied. This usually applies to \acp{TSO}/\acp{ISO} which are responsible for the integrity of the overall grid. \acp{DSO}, however, have no responsibility of overall system balance. For example, they always assume sufficient reserve is available from the upper-level grid, passing the costs of these reserves to their end users. In this case, the chance-constrained optimization approach poses as a more useful tool, minimizing the expected costs while respecting system constraints probabilistically, i.e. with some tolerance of violation (typically 1-10\%). Optimization under uncertainty is challenging because propagating uncertainty through the nonlinear AC power flow equations is computationally hard. In particular, when convexity is needed to guarantee optimality and calculate shadow prices. \cite{bienstock_chance-constrained_nodate} shows that under a linear grid model and Gaussian uncertainty, an exact reformulation of the \ac{CC-OPF} is possible. However, assuming uncertainties to be only Gaussian is severely restrictive as many processes in power grids are better modeled with other distributions \cite{anvari_short_2016}. 

% Dynamic tariffs for end-prosumers are usually based on spot-market clearing prices. These prices reflect the cost associated with power flows on the transmission level, but does not account for cost of local generation and distribution. This means that price signals for end-prosumers are not accurate in providing the true cost of the served energy, hindering optimal investment decisions. In addition, since these prices are determined in the day-ahead market, there is little incentive to adjust consumption based on real-time power balance. Knowing that generation forecast errors for stochastic renewables can be large, the price signals to consumers can be far from reflecting the true cost of the energy. Real-time pricing through intraday and balancing markets are not accessible to end-prosumers and therefore local flexibility can not participate in such markets.

\subsection{Paper's aim}
In this paper we propose an optimization-based \ac{LEM} for accurately pricing active and reactive powers and simultaneously integrate uncertainty of production and consumption through a polynomial-chaos based \ac{CC-OPF}. The use of the \ac{gPC} methodology allows to propagate uncertainty from any stochastic source through the optimization model, without requiring linearity or assuming Gaussian distributions. The combination of \ac{gPC} and the lindistflow equations allows the formulation of a convex, second-order cone problem. Convexity is key in power grid models as it guarantees global optimality and calculation of shadow prices using duality theory. Furthermore, the lindistflow model represents a sufficiently adequate representation of distribution grids, since it accounts for non-approximated longitudinal impedance of branches (that may exhibit high R/X ratio) while neglecting branch shunt admittances that play a less important role in medium- and low-voltage grids. Our method allows aggregators, Distribution Utilities and \acp{DSO} to propose a local market framework that naturally integrates day-ahead scheduling and real-time adjustments through passive balancing without the need to actively bid or commit to a position. Using chance-constrained optimization, techniques for the clearing of both day-ahead and real-time market can be integrated directly in the day-ahead stage. At real-time, the realized uncertainty is measured and the prices are calculated simply by evaluating the polynomial chaos expansion. This ensures very low computational complexity and essentially no need to solve optimization-problems in a time-critical manner within the operational stage. In addition, the grid operator can verify grid feasibility in the day-ahead stage allowing ample time to take remedial action, if necessary. The choice of a passive market, i.e., end-prosumers are not required to submit bids, significantly simplifies the infrastructure requirements. Price signals can be communicated on a common platform and SMs are sufficient to compute the net position of participants and ensure correct billing. The contributions are as follows:

\begin{enumerate}
    \item Formulation of the polynomial-chaos based chance-constrained optimal power flow using the lindistflow approximation allowing accurate modeling of the grid constraints while preserving convexity of the optimization problem.
    \item Using duality theory to extract nodal \acp{PLMP} to establish a two-stage electricity market for day-ahead and real-time clearing.
    \item Analyzing potential cost-savings through end-prosumer participation in the local real-time market through simulations.
    \item Propose a local market framework that ensures satisfaction of grid constraints irrespective of the policy of the market players.
\end{enumerate}

The remainder of the paper is divided as follows: in Section 2 we present the proposed method and give examples of possible strategies for passive market-participants. Section 3 contains four case studies highlighting the performance of the method and scalability. Section 4 provides an in-depth discussion of the results and concludes the paper.

\section{Methods}\label{sec:method}
In this section we detail the mathematical description of our model, including the grid model, polynomial chaos CC-OPF and resulting \acp{PLMP}. We also detail the two proposed flexible prosumer strategies, rule-based and \ac{DP}-based.

\subsection{The lindistflow model} \label{sec:lindistflow}
We consider a radial distribution grid whose topology is expressed as a graph $\mathcal{G} = (\mathcal{N}, \mathcal{L})$ where $\mathcal{N}$ is the set of buses and $\mathcal{L}$ is the set of branches. Due to its radiality, we have: $|\mathcal{N}| = N+1$ and $|\mathcal{L}| = N$. The reduced branch-bus incidence matrix $A\in\mathbb{R}^{N\times N}$ has entries: 
\[
A_{ij} = 
\begin{cases}
+1 & \text{if branch } i \text{ leaves bus } j, \\
-1 & \text{if branch } i \text{ enters bus } j, \\
0 & \text{otherwise}.
\end{cases}
\]
$\mathbf{r}\in\mathbb{R}^N$ and $\mathbf{x}\in\mathbb{R}^N$ denote the vectors of branch resistance and reactance, respectively, while $\mathbf{P}\in\mathbb{R}^N$, $\mathbf{Q}\in\mathbb{R}^N$, $\mathbf{p}\in\mathbb{R}^N$ and $\mathbf{q}\in\mathbb{R}^N$ are the branch active and reactive power and bus active and reactive injections, respectively\footnote{We neglect shunt elements.}. The slack bus does not host DG or loads. The standard lindistflow equations can therefore be written in compact form \cite{kekatos_lecture_nodate}:

\begin{subequations} \label{eq:lindistflow-OPF-original}
\begin{align}
\min_{\mathbf{p}^g, \mathbf{q}^g, P^0, Q^0, \mathbf{V}} \quad & \mathcal{J}(\mathbf{p}^{DG}, \mathbf{q}^{DG} ,P^0, Q^0) \\
        & A^T\mathbf{P} = \mathbf{p} \\
        & A^T\mathbf{Q} = \mathbf{q} \\
        & \mathbf{V} = v_0\mathbf{1} + 2R\mathbf{p} + 2X\mathbf{q} \\
        & \mathbf{p} = \mathbf{p}^{DG} - \mathbf{p}^{d} \\
        & \mathbf{q} = \mathbf{q}^{DG} - \mathbf{q}^{d} \\
        & \underline{V} \leq V \leq \overline{V} \\
        & \norm{\mathbf{p},\mathbf{q}}_2 \leq \mathbf{\overline{f}} \label{eq:lindistflow-OPF-original-linelimit}
\end{align}
\end{subequations}

where $R = FD_rF^T$, $X=FD_xF^T$, $F=A^{-1}$, $D_r=diag(\mathbf{r})$, $D_x=diag(\mathbf{x})$, $\mathbf{\overline{f}}$ is the branch flow limits expressed in terms of power\footnote{Given a branch ampacity limit $I_{ij}^{max}$, we compute $\Bar{f}_{ij} = V^{min}I_{ij}^{max}$, where $V^{min}$ is the minimum allowed operational voltage. This definition of $\Bar{f}_{ij}$ is conservative.}. The injections $\mathbf{p}$ and $\mathbf{q}$ are either \ac{DG} or demand (d): $\mathbf{p} = \mathbf{p}^{DG} - \mathbf{p}^{d}$ and $\mathbf{q} = \mathbf{q}^{DG} - \mathbf{q}^{d}$. Furthermore, the \acp{DG} can be either controllable (e.g. \ac{ESS}) or uncontrollable (e.g. \ac{PV}). We consider a generic, convex objective $\mathcal{J}$, representing power injections from \acp{DG} and exchanges from the slack bus. $\underline{V}$ and $\overline{V}$ are the squared voltage lower and upper bounds, respectively. 

%\textcolor{red}{Note here the difference of the voltage equation. this formulation uses the injection power, not the branch flow as in the code.}

\subsection{Chance-constrained Polynomial-Chaos OPF}
gPC allows to propagate uncertainty, from input random variables to output random variables, through a complex model.
In this work, we use an intrusive polynomial chaos approach, where deterministic equations are projected onto a polynomial basis \cite{sullivan_introduction_2015}. To fix the ideas, we consider a random vector, called the \textit{stochastic germ}, $\boldsymbol{\xi} = [\xi_1,...,\xi_{d_\xi}]$ where every element is independent and with finite variance. We can approximate any random variable with finite variance as a function of orthogonal polynomials and coefficients, i.e. $X=\sum_{i=0}^{K-1} x_i\Psi_i(\boldsymbol{\xi})$, where $K=\frac{(p_\xi+d_\xi)!}{p_\xi!d_\xi!}$, $p_\xi$ is the polynomial degree and $d_\xi$ is the number of elements in the stochastic germ. $\Psi_i$ is the i-th polynomial. For more details on intrusive polynomial chaos applications in power systems, the reader is directed to \cite{muhlpfordt_generalized_2018, muhlpfordt_chance-constrained_2019}.

First, we apply the Galerkin-projection on the linear equations in \cref{eq:lindistflow-OPF-original} which results in $K$-times the number of equations. We denote by subscript $k \in\{0,...,K-1\}$, the $k$-th PC-coefficient for any variable, i.e. $\mathbf{p}_k$ refers to the vector of active power injections for the k-th PC-coefficient. Any individual bus is denoted by superscript $n$, i.e. $p_k^n$ refers to the $k$-th PC-coefficient of the active power injection in bus $n$. Finally, we also consider the subscript $t$ to represent the timestep. Certain properties of polynomial chaos are useful to formulate the OPF-problem. Notably, $E[X] = x_0$ and $Var[X] = \sum\limits_{i=1}^{K-1}x_i^2$, i.e. the mean of the random variable is simply the 0-th PC-coefficient, and the variance is the sum of the square of the PC-coefficients, excluding the 0-th coefficient. 

Using the aforementioned definitions, we can formulate the following chance-constrained optimization problem:

\begin{equation}
\min_{\{\mathbf{p}_k^{DG}, \mathbf{q}_k^{DG}\}_{k=0}^{K-1}} 
\quad 
\sum_{t\in\mathcal{T}}\mathcal{J}(\mathbf{p}_{0,t}^{DG}, \dots, \mathbf{p}_{K-1,t}^{DG}, P_{0,t}^0, \dots, P_{K-1,t}^0)
\label{eq:ccopf-obj}
\end{equation}

% equality-constraints
% real power balance
\begin{equation}
A^{\top} \mathbf{P}_{k,t} = \mathbf{p}_{k,t}, : (\boldsymbol{\lambda}_{k,t}), 
\quad \forall k \in \mathcal{K}, \forall t \in \mathcal{T}
\label{eq:ccopf-b}
\end{equation}

% reactive power balance
\begin{equation}
A^{\top} \mathbf{Q}_{k,t} = \mathbf{q}_{k,t}, : (\boldsymbol{\mu}_{k,t}), 
\quad \forall k \in \mathcal{K}, \forall t \in \mathcal{T}
\label{eq:ccopf-c}
\end{equation}

% voltage balance
\begin{equation}
\mathbf{V}_{k,t} = v_0 \mathbf{1} + 2R\mathbf{p}_{k,t} + 2X\mathbf{q}_{k,t}, 
\quad \forall k \in \mathcal{K}, \forall t \in \mathcal{T}
\label{eq:ccopf-d}
\end{equation}

% Injection balance
\begin{equation}
\mathbf{p}_{k,t} = \mathbf{p}_{k,t}^{DG} - \mathbf{p}_{k,t}^d, 
\quad \forall k \in \mathcal{K}, \forall t \in \mathcal{T}
\label{eq:ccopf-e}
\end{equation}

% inequality-constraints
% Active power
\begin{equation}
\norm{ \biggl(P_{1,t}^l,...,P_{K-1,t}^l\biggr) }_2
 \leq  \frac{\overline{f}_l-P_{0,t}^l}{\sqrt{2}\Gamma(\epsilon)}, 
\quad \forall l \in \mathcal{L}, \forall t \in \mathcal{T}
\label{eq:ccopf-f}
\end{equation}

% Reactive power
\begin{equation}
\norm{ \biggl(Q_{1,t}^l,...,Q_{K-1,t}^l\biggr) }_2
 \leq  \frac{\overline{f}_l-Q_{0,t}^l}{\sqrt{2}\Gamma(\epsilon)}, 
\quad \forall l \in \mathcal{L}, \forall t \in \mathcal{T}
\label{eq:ccopf-g}
\end{equation}

% Voltage upper bound
\begin{equation}
\norm{ \biggl(V_{1,t}^n,...,V_{K-1,t}^n\biggr) }_2
 \leq  \frac{\overline{V} - V_{0,t}^n}{\Gamma(\epsilon)}, 
\quad \forall n \in \mathcal{N}, \forall t \in \mathcal{T}
\label{eq:ccopf-h}
\end{equation}

% Voltage lower bound
\begin{equation}
\norm{ \biggl(V_{1,t}^n,...,V_{K-1,t}^n\biggr) }_2
 \leq  \frac{V_{0,t}^n-\underline{V}}{\Gamma(\epsilon)}, 
\quad \forall n \in \mathcal{N}, \forall t \in \mathcal{T}
\label{eq:ccopf-i}
\end{equation}

% Distributed generation upper bound
\begin{equation}
\norm{ \biggl(p_{1,t}^{DG,n},...,p_{K-1,t}^{DG,n}\biggr) }_2
 \leq  \frac{\overline{p^{DG,n}} - p_{0,t}^{DG,n}}{\Gamma(\epsilon)}, 
\quad \forall n \in \mathcal{N}, \forall t \in \mathcal{T}
\label{eq:ccopf-j}
\end{equation}

% Distributed generation lower bound
\begin{equation}
\norm{ \biggl(p_{1,t}^{DG,n},...,p_{K-1,t}^{DG,n}\biggr) }_2
 \leq  \frac{p_{0,t}^{DG,n}-\underline{p^{DG,n}}}{\Gamma(\epsilon)}, 
\quad \forall n \in \mathcal{N}, \forall t \in \mathcal{T}
\label{eq:ccopf-k}
\end{equation}

where the objective is to minimize the cost of local generation ($\mathbf{p}^g$) and the slack injection $P^0$: 
\begin{equation}
\begin{split}
\mathcal{J}(\mathbf{p}_0^g, \dots, \mathbf{p}_{K-1}^g, P_0^0, \dots, P_{K-1}^0)
&= {}\\[1em]
&\hspace{-10em}\left(\mathbf{c}^{local}\right)^T\mathbf{p}_0^{g} 
+ \mathbf{p}_0^{g} C_1^{local} \left(\mathbf{p}_0^{g}\right)^T \\
&\hspace{-10em}+ \sum_{k=1}^{K-1} \mathbf{p}_k^{g} C_2^{local} \left(\mathbf{p}_k^{g}\right)^T 
+ c^{slack}P_0^{0} + C^{slack}\left(P_0^{0}\right)^2  \\
&\hspace{-10em}+ \sum_{k=1}^{K-1} C_2^{slack}\left(P_k^{0}\right)^2.
\end{split}
\end{equation}

The rationale from the chosen cost-function is that in a standard CC-OPF setting, we are optimizing $E[\mathcal{J}(\boldsymbol{X})]$, where $\mathcal{J}$ is a convex-quadratic function in the random vector $\boldsymbol{X}$, i.e. $\mathcal{J}(\boldsymbol{X}) = a^TX+\boldsymbol{X}^TB\boldsymbol{X}$. We therefore obtain: $E[\mathcal{J}(\boldsymbol{X})] = a^TE[\boldsymbol{X}] + tr(BV[\boldsymbol{X}]) + E[\boldsymbol{X}]^TBE[\boldsymbol{X}]$, where the expectance and variance can be written in terms of the PC-coefficients. 

Constraints \cref{eq:ccopf-b,eq:ccopf-c,eq:ccopf-d,eq:ccopf-e} are the lindistflow equations written in terms of the \ac{gPC}-expansion. Constraints \cref{eq:ccopf-f,eq:ccopf-g} are branch power flow constraints, \cref{eq:ccopf-h,eq:ccopf-i} are squared voltage magnitude constraints and \cref{eq:ccopf-j,eq:ccopf-k} are resource-constraints, all reformulated as chance-constraints. $\Gamma(\epsilon)$ represents the parameter that adjusts the risk-level $\epsilon$, i.e. for a random variable $X$, $\text{Pr}(X\leq0) \geq 1-\epsilon$\footnote{For example: in the case of a Gaussian distribution, we have $\Gamma(\epsilon) = \Phi^{-1}(1-\epsilon)$, where $\Phi$ is the Gaussian distribution function. A distributionally robust bound can also be considered \cite{kuhn_distributionally_2025}.}.

To preserve convexity, we consider inner box constraint approximations for the branch flows in \cref{eq:ccopf-f,eq:ccopf-g}. The resulting problem is convex, since the objective is convex quadratic and constraints are linear and second-order cones. 

% \begin{figure}[H]
%     \centering
%     \includegraphics[width=0.45\textwidth]{box-approx.png}
%     \caption{Inner box approximation for branch flow limits.}
%     \label{fig:box-approx}
% \end{figure}

\subsection{Obtaining LMP samples}
Once the model in \cref{eq:ccopf-obj,eq:ccopf-b,eq:ccopf-c,eq:ccopf-d,eq:ccopf-e,eq:ccopf-f,eq:ccopf-g,eq:ccopf-h,eq:ccopf-i,eq:ccopf-j,eq:ccopf-k} has been solved, the PC-coefficients for the duals for the active and reactive power balance, $\boldsymbol{\lambda}_{k,t}\in\mathbb{R}^N$ and $\boldsymbol{\mu}_{k,t}\in\mathbb{R}^N$, can be obtained. These PC-coefficients, together with the stochastic germ defines the \acp{PLMP}. Since the 0'th PC-coefficient equals the mean value, this also defines the deterministic day-ahead price; $\pi_{DA,t}^n=\lambda_{0,t}^n$. By measuring the realization of the stochastic germ $\boldsymbol{\xi}$, and then evaluating the PC-expansion, we can obtain a realization of the \acp{PLMP}, which we call the realtime price; $\pi_{RT,t}^n$. For ease of notation, we will use only $\pi_{DA}$ and $\pi_{RT}$ to denote a generic day-ahead and realtime price where the specific bus is implied. 

In general, for any stochastic variable in the problem, we can obtain its realization by measuring the stochastic germ and evaluating the polynomial expansion. For example, consider an output random variable $X$ with PC-expansion: $X = \sum_{i=0}^{K-1} x_i\Psi_i(\boldsymbol{\xi})$. After solving the model to obtain the PC-coefficients $x_i$ of $X$, we measure the germ $\boldsymbol{\xi} = [\xi_1,...,\xi_d]$ and can directly estimate $X$. To express the full probability distribution, for the day-ahead stage, we sample a large number of i.i.d. samples from the germ and evaluate the polynomial expansion for every stochastic variable.

\subsection{Interaction with Central Electricity Markets}
The \ac{CEM} day-ahead market is cleared separately and before the \ac{LEM}, producing a deterministic day-ahead price for every delivery period of the next day. The day-ahead clearing implicitly includes the price-elasticity of the end-prosumer demand, meaning that day-ahead arbitrage has already been accounted for. This is equivalent to the current principle of day-ahead spot markets. We therefore only model the realtime arbitrage opportunities by end-prosumers, i.e. they adapt their strategy based on the delta-price, the difference between the realtime and day-ahead price: $\pi_\Delta = \pi_{RT}-\pi_{DA}$. Rational end-prosumers should increase their consumption (resp. decrease their production) when the delta-price is negative and reduce their consumption (resp. increase their production) when it is positive.
\noindent
\subsection{Flexible prosumer strategies}
As discussed, \acp{LEM} are challenging to implement due to the large number of end-prosumers. To simplify the operation of the \ac{LEM} we propose to treat end-prosumers as passive market participants. Since day-ahead schedules have been implicitly determined, the end-prosumers can perform arbitrage based on the difference between the realtime and day-ahead prices. In this respect, we consider two types of rational, profit-seeking end-prosumers that can freely choose their strategy\footnote{The prosumers' local agents can even include forecasts that outperform that of the \ac{LEM} \ac{MO} to maximize their own revenue.}. The first is rule-based, using only the delta-price for the current hour to decide actions. This agent never incurs losses, as its actions are always optimal at the current timestep. However, since it has limited storage capacity, the State Of Charge \ac{SOC} can become saturated, preventing the agent from taking some actions. For example, if the SOC is 0\%, the agent can not discharge and therefore can not profit if the realtime price-delta is positive. The second agent improves on this by also using the probabilistic forecasts of \acp{LMP} to better plan resource-utilization. Its implementation is based on a dynamic programming framework where the agent computes an optimal trajectory starting from a terminal condition. Applying only the first timestep, the agent acts in a receding horizon fashion. 

For both agents, we consider a limited energy storage capacity $E_{cap}$, limited power capacity $P_{cap}$ and initial and final SOC: $E_{init} = E_{end} = \frac{E_{cap}}{2}$. The market clearing period is hourly ($\Delta t=1$) and the power to energy ratio (C-rating) is 0.25. 

\subsubsection{Rule-based agent}
The rule-based agent acts on the realtime $\pi_{\Delta}$, while respecting energy and power capacity for all hours $t\in\{1,...,22\}$. To ensure that the net energy exchanged remains 0 over the time-horizon, the actions in hours 23 and 24 ensures that it can return to the $E_{end}$ requirement. The rule-based agent can be considered a greedy agent, i.e. it always chooses the optimal strategy for the current time step, without considering future possible realizations. 

\begin{equation}
\begin{aligned}
&\forall t\in \{1,...,22\}: \\
p_t &=
    \begin{cases}
      -\min\left(\frac{E_t}{\Delta t},\,P_{\text{cap}}\right), &
          \pi_{\Delta,t} \geq 0 \quad\text{(discharge)},\\[6pt]
      \phantom{-}\min\left(\frac{E_{\text{cap}} - E_t}{\Delta t},\,P_{\text{cap}}\right), &
          \pi_{\Delta,t} < 0 \quad\text{(charge)},
    \end{cases} \\[8pt]
&t=23:\\
p_{23} &=
  \begin{cases}
    \;\;P_{\text{cap}}, & E_{23} < E_{init} - P_{\text{cap}}\Delta t,\\
    -P_{\text{cap}}, & E_{23} > E_{init} + P_{\text{cap}}\Delta t
  \end{cases} \\
&t=24:\\
p_{24} &= \frac{E_{end} - E_{24}}{\Delta t},
\end{aligned}
\end{equation}
where $p_t$ and $E_t$ are the power setpoint and energy level at time $t$. 

\subsubsection{DP-based agent}
The \ac{DP}-based agent uses a multi-stage stochastic program with perfect recourse. To this end, we consider the value-function $V_t$ at every timestep, calculating the optimal power-setpoint at that timestep. Then, we calculate the optimal recursive trajectory considering a terminal cost:
\begin{equation}
    \begin{aligned}
    V_H(E_H) &=
        \begin{cases}
            0 & \text{if} \ \ E_H = E_{end}\\
            -\kappa & \text{else},
        \end{cases}
    \end{aligned}
\end{equation}
where $\kappa$ is some constant.
We consider discrete sets of states and actions, $\mathcal{S}$ and $\mathcal{P}$ respectively. At every stage, we can formulate the following set of Bellman equations:

\begin{equation}
\begin{aligned}
& t=1: \\
& V_1(E_t) = \max_{p_t\in\mathcal{P}(E_t)} \{p\pi_{\Delta,1} +V_1(E_t+p) \} \\[6pt]
& \forall t\in\{1,...,23\}: \\
& \begin{cases}
V_t^{up}(E_t) &= \max_{p_t^{up}\in\mathcal{P}(E_t)} \{p_t^{up}v_t^{up} + V_{t+1}(E_t+p) \} \\[6pt]
V_t^{down}(E_t) &= \max_{p_t^{down}\in\mathcal{P}(E_t)} \{p_t^{down}v_t^{down} + V_{t+1}(E_t+p) \} \\[6pt]
V_t(E_t) &= q_tV_t^{up}(E_t) + (1-q_t)V_t^{down}(E_t)
\end{cases} \\[6pt]
& t=24: \\
& \begin{cases}
V_{24}^{up}(E_t) &= \max_{p_t^{up}\in\mathcal{P}(E_t)} \{p_{24}^{up}v_t^{up} - V_H(E_H) \} \\[6pt]
V_{24}^{down}(E_t) &= \max_{p_t^{down}\in\mathcal{P}(E_t)} \{p_{24}^{down}v_t^{down} - V_H(E_H) \} \\[6pt]
V_{24}(E_t) &= q_tV_t^{up}(E_t) + (1-q_t)V_t^{down}(E_t) \\[6pt]
\end{cases}
\end{aligned}
\end{equation}

\noindent
Where $q_t=Pr(\pi_{\Delta,t}>0)$, i.e. the probability that the delta-price is larger than 0 for a specific time step. At each timestep, the \ac{DP}-based agent solves recursively problems $V_{24},...,V_t$, and then applies the optimal decision $p_t$. This essentially allows the agent to anticipate periods of high price-deltas and optimally scheduling its state for those periods.

\section{Results}\label{sec:result}
\begin{figure}[H]
    \centering
    \includegraphics[width=0.5\textwidth]{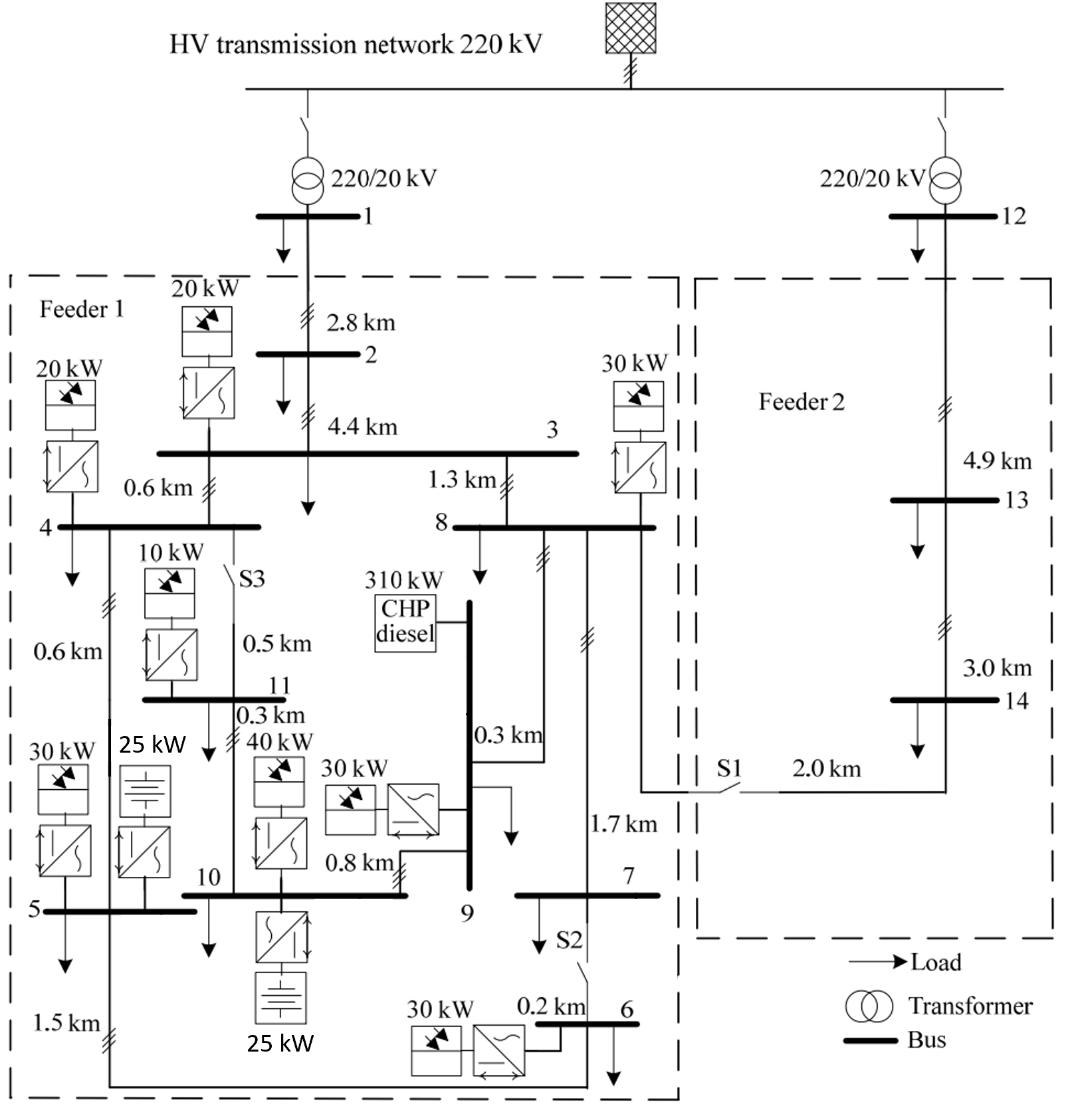}
    \caption{Grid for case-studies 1-3, which is a modified version of the medium-voltage distribution network benchmark developed by the CIGRÉ Task Force C6.04.02 \cite{conseil_international_des_grands_reseaux_electriques_benchmark_2014}.}
    \label{fig:grid}
\end{figure}

We showcase the method across four case-studies, each focusing on a particular feature of the market clearing framework. For the sake of reproducibility, the grid considered for the case studies 1-3 is based on the CIGRÉ Task Force C6.04.02 \cite{conseil_international_des_grands_reseaux_electriques_benchmark_2014}. It is radial and contains 14 buses, visualized in \cref{fig:grid}. The slack bus has index 0 and represents the interface with the upper-level grid. Case-study 1 explores the potential behavior by agents located at different locations in the grid. We show that a rational agent with a flexible asset can optimize its profit while helping the grid to reduce the risk of congestion or voltage violations. Case-study 2 shows a system with congestion and how it impacts the distribution of the \acp{LMP}. Finally, case-study 3 shows a system with active voltage constraints, showing how these create differences between the \acp{LMP} at different buses. We also show an a posteriori analysis of the system state using non-approximated AC load flow calculations, highlighting that the passive balancing by local flexible agents does not exacerbate the voltage-constrained bus. Case-study 4 considers a larger, 179-bus system, based on the Oberrhein \ac{MV} network \cite{pandapower_mv_nodate}, where our goal is to highlight the computational scalability of the proposed method. 

All case studies consider typical daily load profiles, generated from the dataset provided in \cite{lorenzo_hierarchical_2019}. PV profiles are generated synthetically (see example in \cref{fig:case135-config}).
Grid models and full details of case study parameters can be found in the Supplementary Data. A summary is reported in \cref{tab:case_params}.

\subsection{Case I: Consumer reaction to real-time pricing}
\begin{figure}[H]
    \centering

    % Top row
    \begin{subfigure}[t]{0.24\textwidth}
        \centering
        \includegraphics[width=\textwidth]{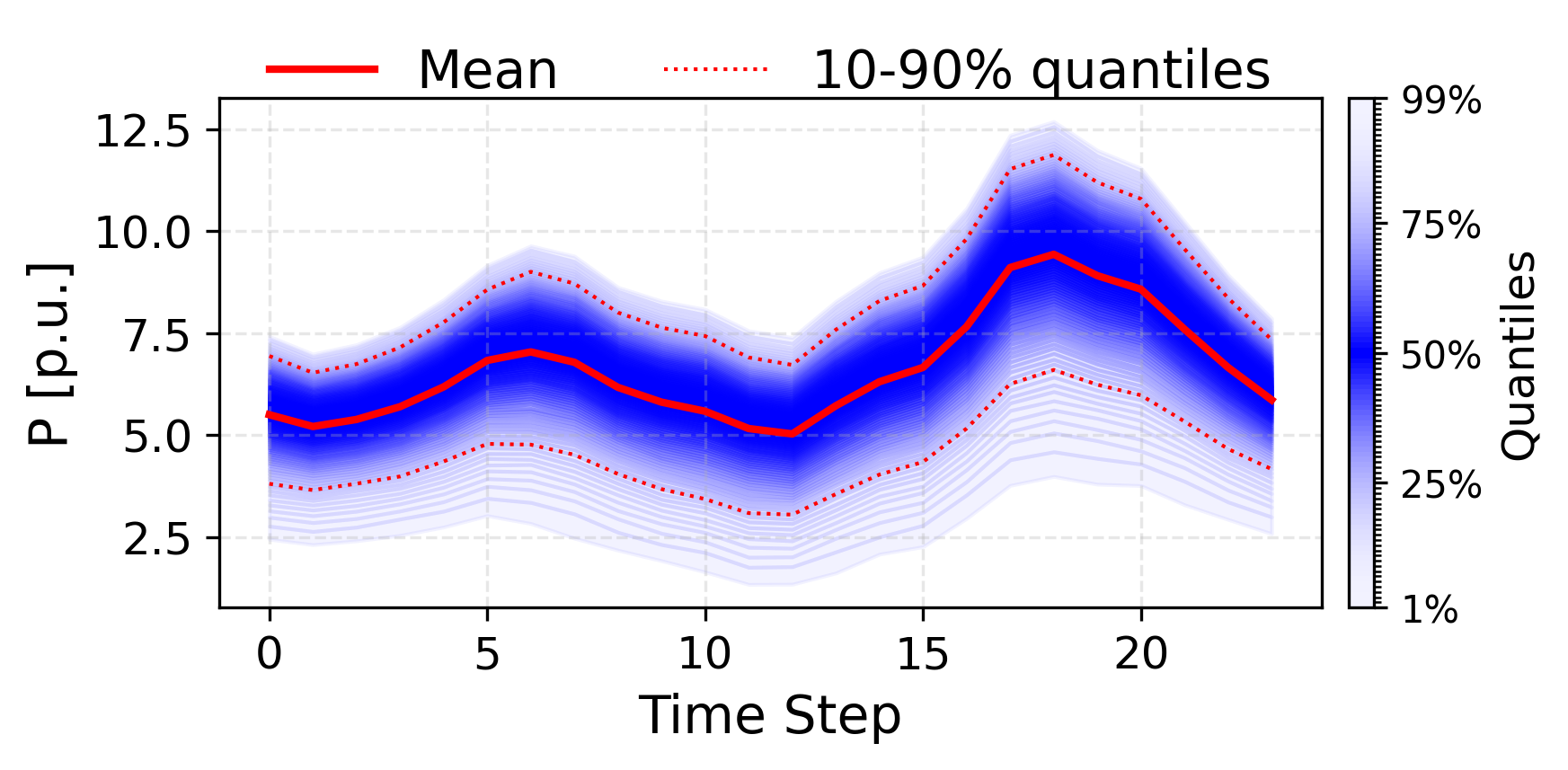}
        \caption{Distribution of real power flow at the slack bus.}
        \label{fig:case1-slack-p}
    \end{subfigure}
    \hfill
    \begin{subfigure}[t]{0.24\textwidth}
        \centering
        \includegraphics[width=\textwidth]{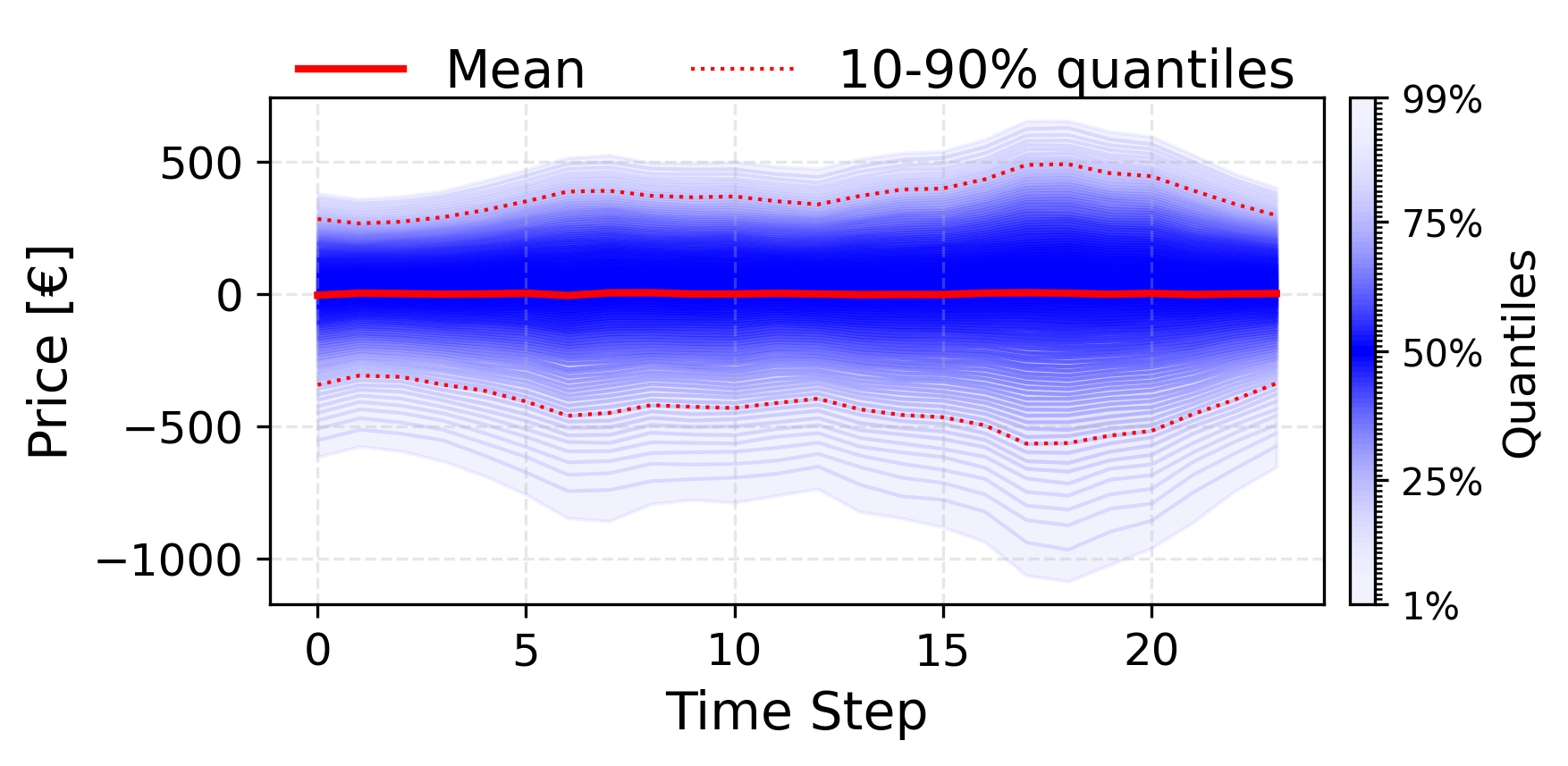}
        \caption{Distribution of $\pi_\Delta$ at bus 10.}
        \label{fig:case1-lmp-ess-node}
    \end{subfigure}

    \vspace{0.5cm} % Space between rows

    % Bottom row
    \begin{subfigure}[t]{0.24\textwidth}
        \centering
        \includegraphics[width=\textwidth]{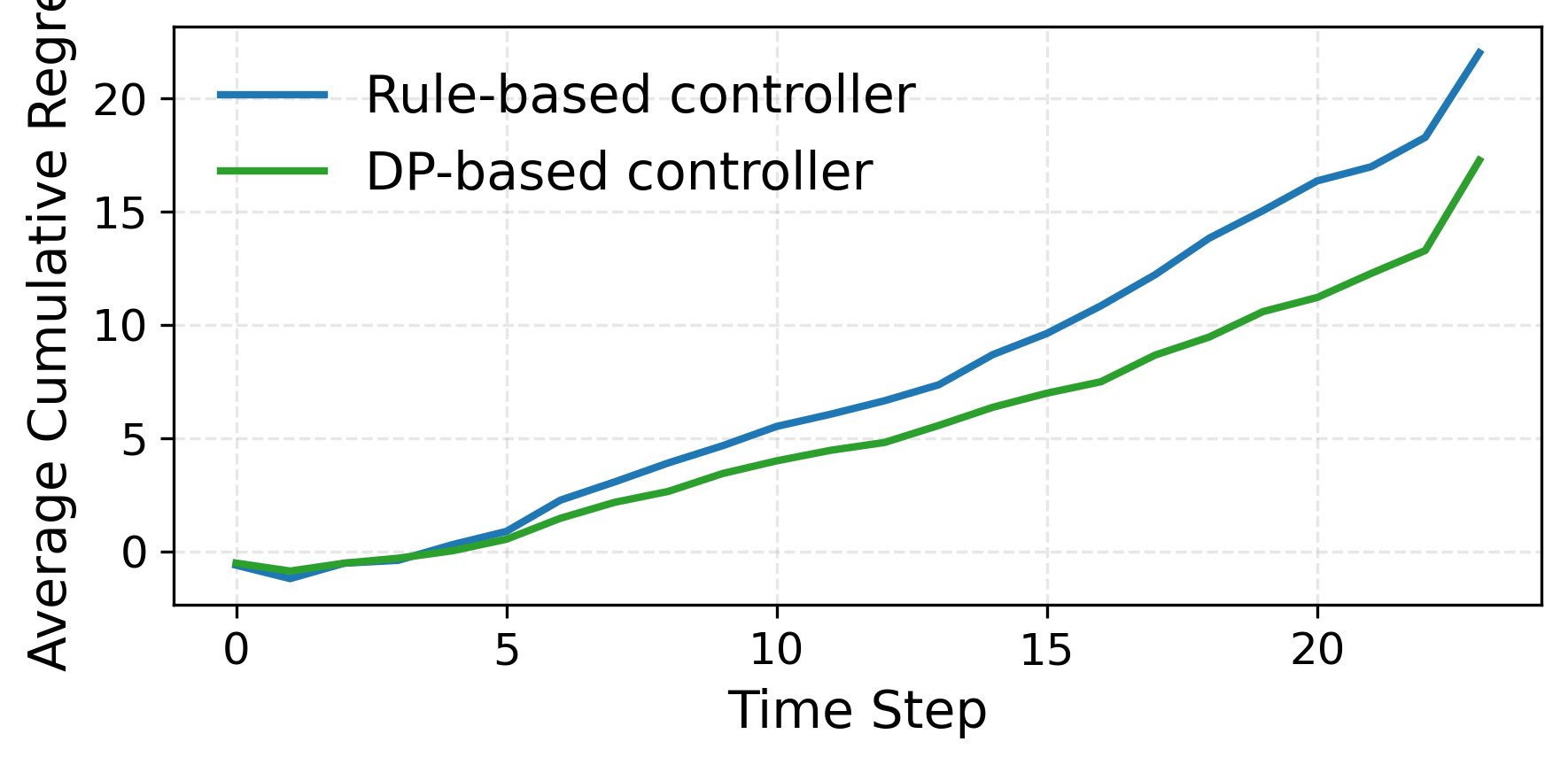}
        \caption{Average cumulative regret for rule-based and \ac{DP}-based controllers, with respect to the in-hindsight controller.}
        \label{fig:case1-regret}
    \end{subfigure}
    \hfill
    \begin{subfigure}[t]{0.24\textwidth}
        \centering
        \includegraphics[width=\textwidth]{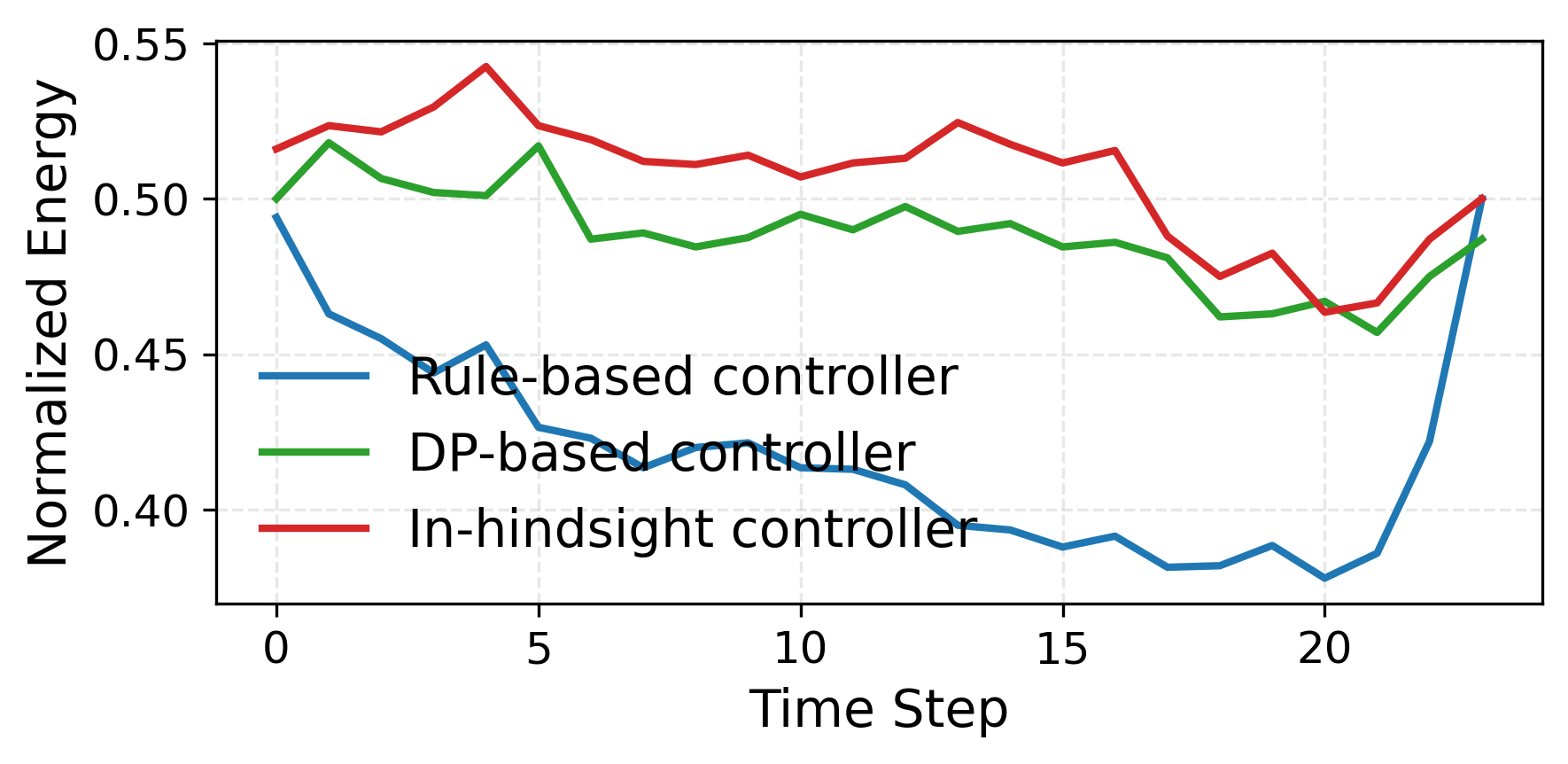}
        \caption{Average SOC across the time-horizon for the rule-based, \ac{DP}-based and in-hindsight controller normalized between [0,1].}
        \label{fig:case1-soc}
    \end{subfigure}

    \caption{Upper left and right: power flow schedule at the slack bus and arbitrage price seen by the \ac{ESS} in bus 10. Lower left and right: average cumulative regret and average \ac{ESS} Energy level across scenarios.}
    \label{fig:ess_summary}
\end{figure}

In this case-study we focus on the real-time pricing mechanism, intended to encourage passive balancing among end-prosumers to reduce the dispatch-error of the DSO. In the intraday period, the \ac{MO} issues real-time prices, as a result of measuring the realized uncertainty in the local distribution system it operates. As discussed, end-prosumers perform arbitrage based on the price delta, $\pi_\Delta$. To illustrate the proposed methodology, we consider end-prosumers in buses 5 and 10 with access to an \ac{ESS} (c.f. \cref{fig:grid}), which can, for example, be an asset equipped with a \ac{HEMS}. \cref{fig:case1-slack-p,fig:case1-lmp-ess-node} show the distribution of the power flow in the slack bus and the \ac{LMP} in bus 10, respectively. The red line represents the expected values, i.e. the day-ahead clearing, while the shades of blue represent the uncertainty. Notably, the distribution of the \ac{LMP} in the \ac{ESS} bus has a period of high volatility in the afternoon. 

The policy implemented by the end-prosumer can be of arbitrary complexity, but here we show two typical examples. The first example considers a simple rule-based control where the \ac{ESS} consumes when $\pi_\Delta <0$ and produces when $\pi_\Delta>0$. The second example considers a more complex control based on \ac{DP}. This latter prosumer's controller calculates the best action by estimating profits backwards from a terminal condition. It uses the information in the probability distribution of the realtime prices calculated in the day-ahead stage and can, therefore, anticipate periods of higher price volatility and reserve capacity for those events. The performance of the controllers is measured in regret with respect to the in-hindsight controller; a controller with perfect foresight of realtime prices, shown in \cref{fig:case1-regret}. We observe that the \ac{DP}-based controller has lower regret, highlighting its capacity to account for the probabilistic forecasts of \acp{LMP}. Indeed, \cref{fig:case1-soc} show the \ac{SOC} of the \ac{ESS} for the three controllers. It can be seen that the \ac{DP}-based controller anticipates the higher price-uncertainty by reserving capacity, similarly to the in-hindsight controller.

\subsection{Case II: Congestion pricing}

\begin{figure}[H]
    \centering

    % Top row
    \begin{subfigure}[t]{0.24\textwidth}
        \centering
        \includegraphics[width=\textwidth]{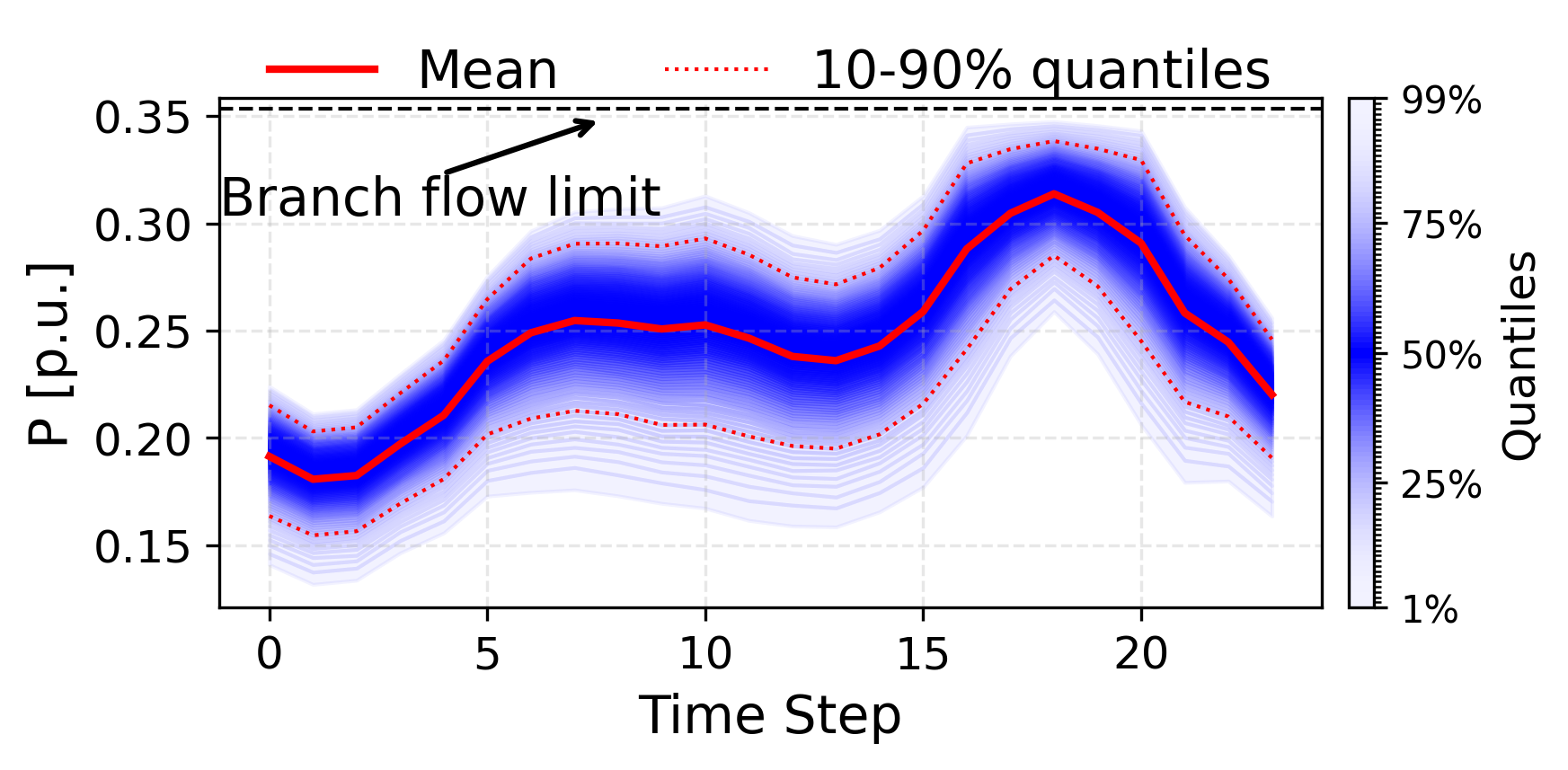}
        \caption{Distribution of branch real power flow in line 8-9, the congested line.}
        \label{fig:case2_p_flow}
    \end{subfigure}
    \hfill
    \begin{subfigure}[t]{0.24\textwidth}
        \centering
        \includegraphics[width=\textwidth]{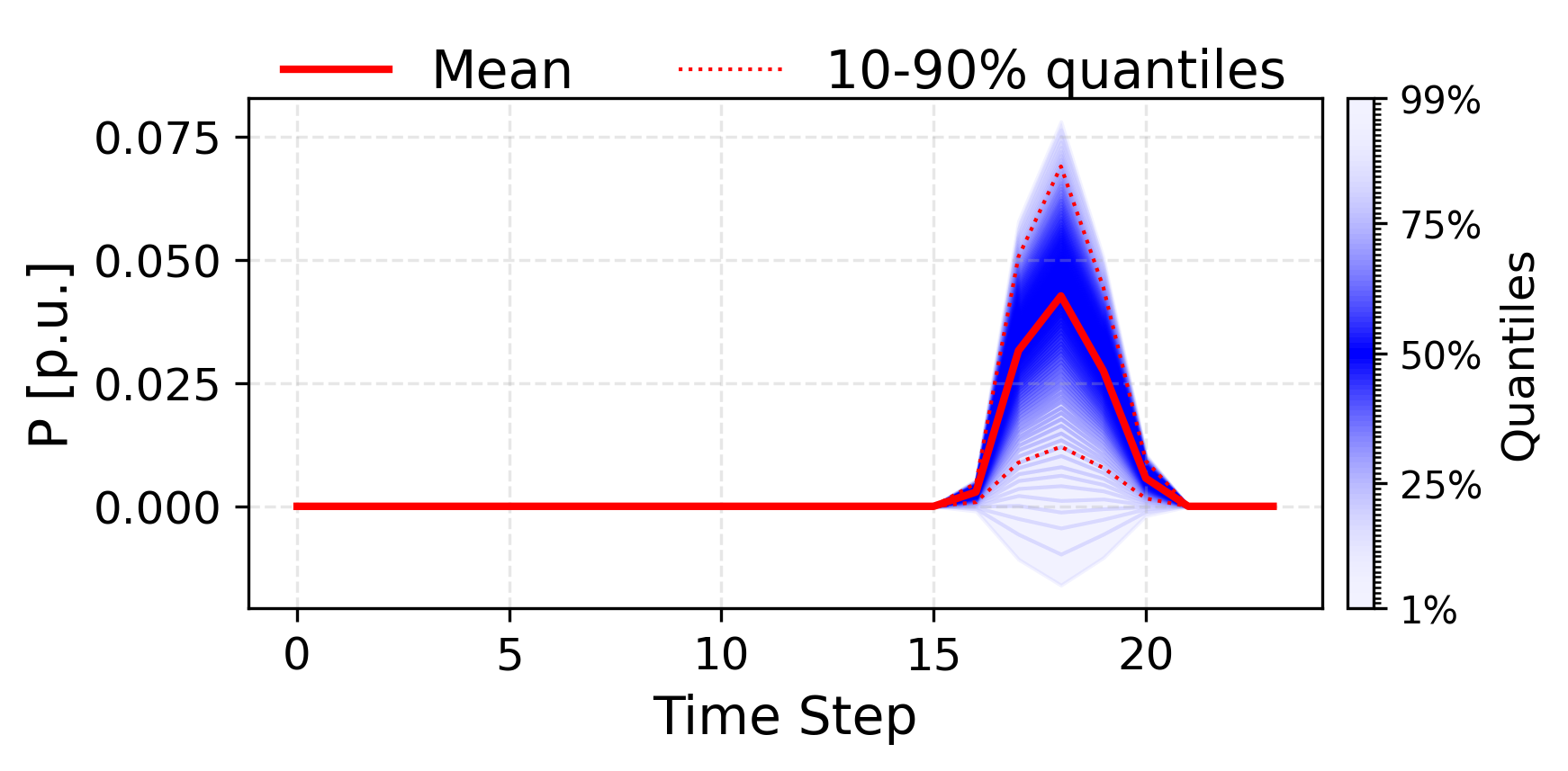}
        \caption{Distribution of power injected at bus 9 by the CHP Diesel generator.}
        \label{fig:case2_pinj_generator}
    \end{subfigure}

    %Bottom row
    \begin{subfigure}[t]{0.24\textwidth}
        \centering
        \includegraphics[width=\textwidth]{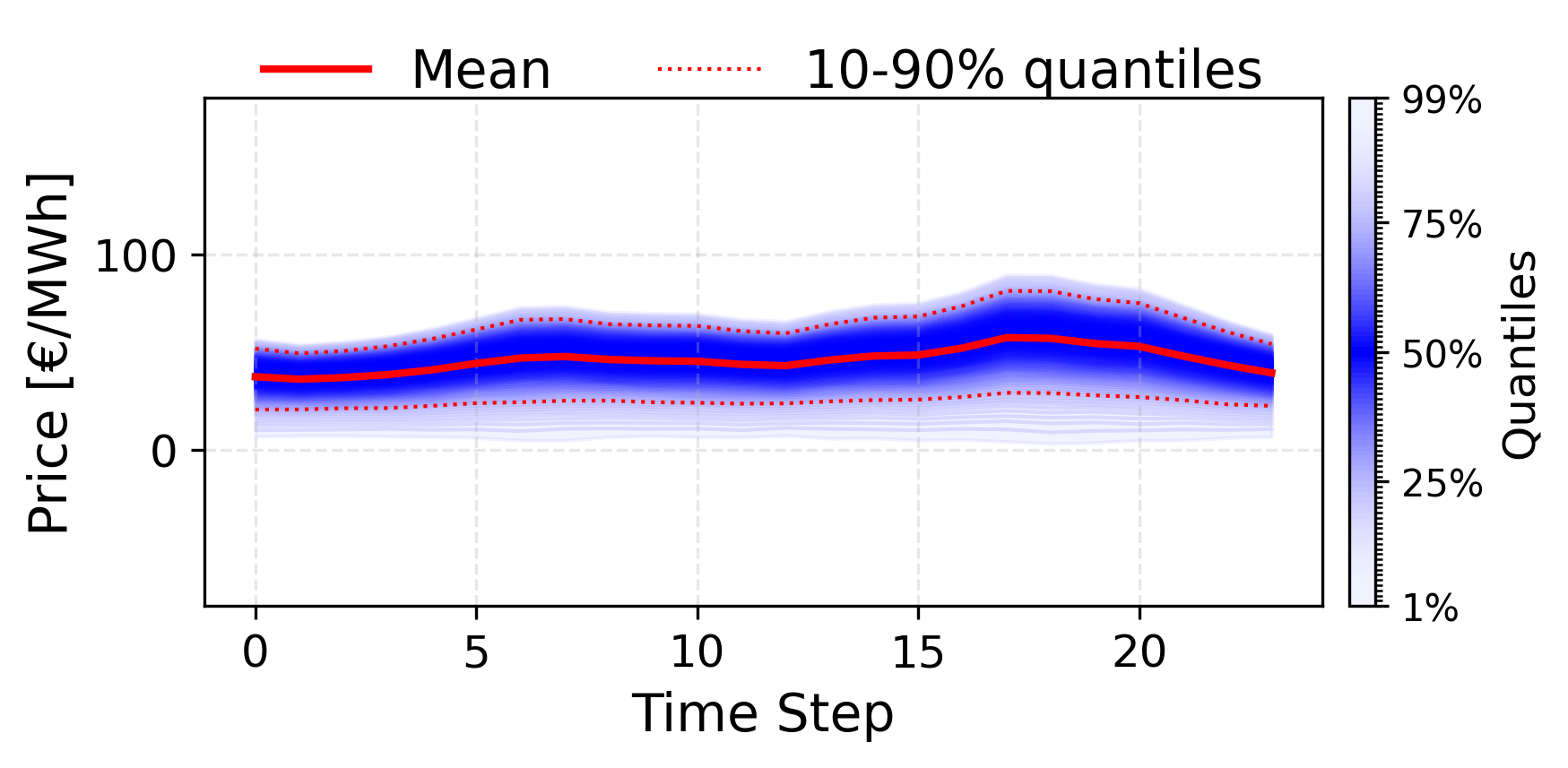}
        \caption{Distribution of \acp{PLMP} in bus 1, which is in the uncongested part of the grid.}
        \label{fig:case2_lmp_p_uncongested}
    \end{subfigure}
    \hfill
    \begin{subfigure}[t]{0.24\textwidth}
        \centering
        \includegraphics[width=\textwidth]{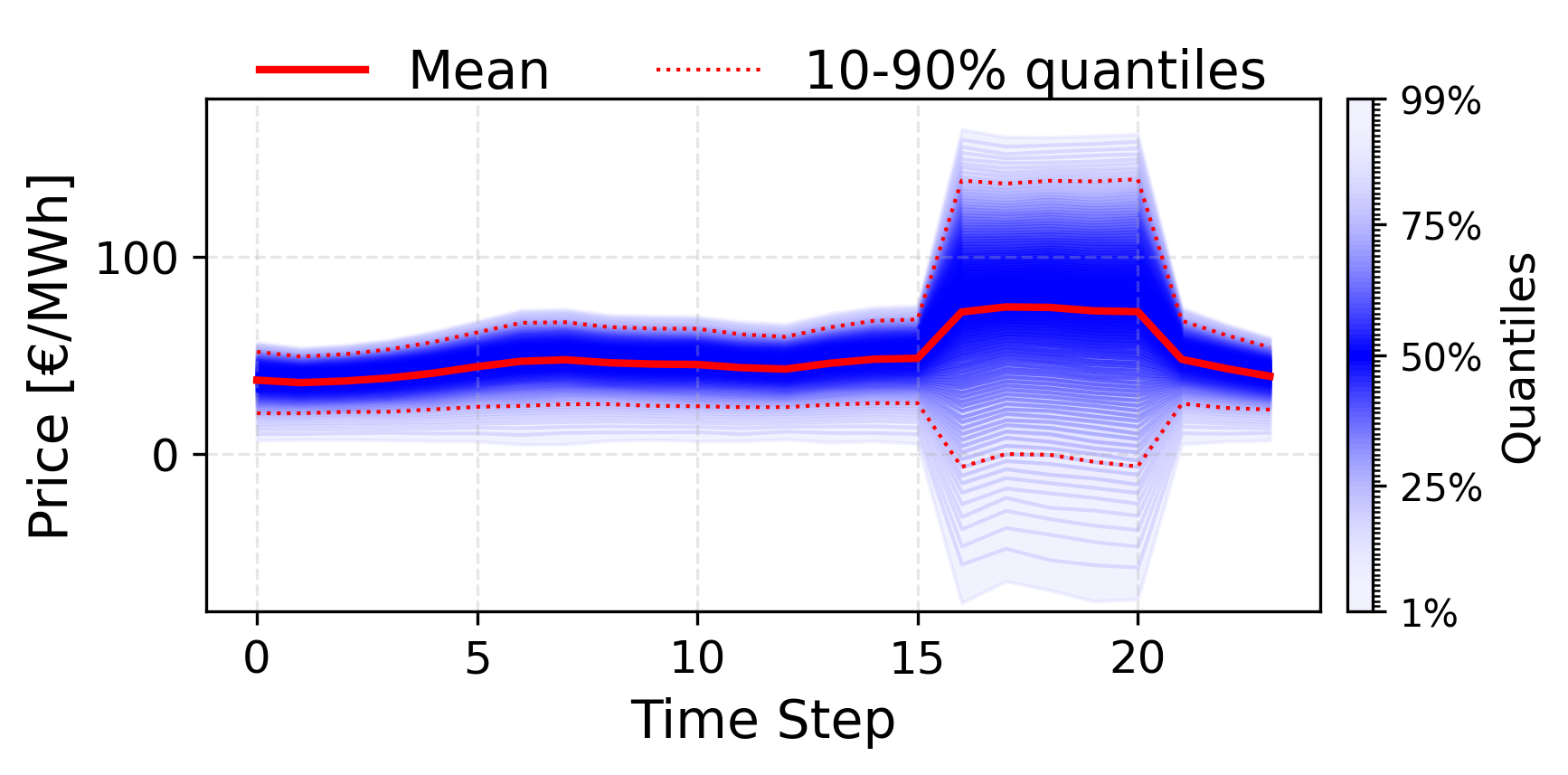}
        \caption{Distribution of \acp{PLMP} in bus 9, which is in the congested part of the grid.}
        \label{fig:case2_lmp_p_congested}
    \end{subfigure}
    \caption{Case study on local congestion and resulting \acp{PLMP}. The congestion occurs at timestep 10, when the load in node 4 increases. The resulting \acp{PLMP} show a congested system.}
    \label{fig:case2}
\end{figure}

In the second case-study we explore how binding network constraints impact the spatial distribution of prices. In this respect, the capacity of the branch between buses 8 and 9 (c.f. \cref{fig:grid}) has been artificially reduced to provoke a congestion, resulting in reduced operational envelope for the CHP diesel generator in bus 9 as well as loads and \ac{PV}-plants in buses 9, 10 and 11. We fix the cost parameters of the CHP diesel generator to be larger than the cost of supplying power from the upper level grid, resulting in a preference of the latter. As is observed in \cref{fig:case2_p_flow}, when the load in the system increases, the power flow in branch 8-9 reaches its limit, creating a congestion. The rest of the demand in buses 9, 10 and 11 must therefore be served by the local CHP diesel generator in bus 9 (\cref{fig:case2_pinj_generator}).

This is reflected in the \acp{PLMP} (ref. \cref{fig:case2_lmp_p_congested,fig:case2_lmp_p_uncongested}), where the distribution of \acp{PLMP} vary between the buses in the congested and uncongested areas. Notably, when there is congestion, the \acp{PLMP} vary between the two sides of the congestion effectively showing the congestion rent on that line. Since the market clearing takes into account uncertainty from producers and prosumers, we also observe the effect of uncertainty on the \acp{PLMP}. For example: when the network becomes congested, any uncertainty in the part of the grid which is congested is covered by the flexibility of the local generator. This is observed in \cref{fig:case2_lmp_p_congested}, where the spread of the \acp{PLMP} increases when the congestion occurs.

% \begin{figure}[H]
%     \centering
%     \includegraphics[width=0.9\textwidth]{case2_lmp_p_samples.png}
%     \caption{Distributions of real power LMPs.}
%     \label{fig:cs2}
% \end{figure}

\subsection{Case III: Voltage constrained system}
In this case we focus on voltage violations, which are particularly important in distribution grids. In this respect, the lindistflow formulation accounts for nodal voltages and we consider that the nodal voltage magnitudes must lie between 0.95 and 1.05 p.u. To simulate the effect of active voltage constraints, we artificially increase the length of the branch between buses 8 and 9 (c.f. \cref{fig:grid}). We also consider a significant increase in the installed \ac{PV} capacity in bus 9.

\begin{figure*}[t]
    \centering
    \includegraphics[width=0.95\textwidth]{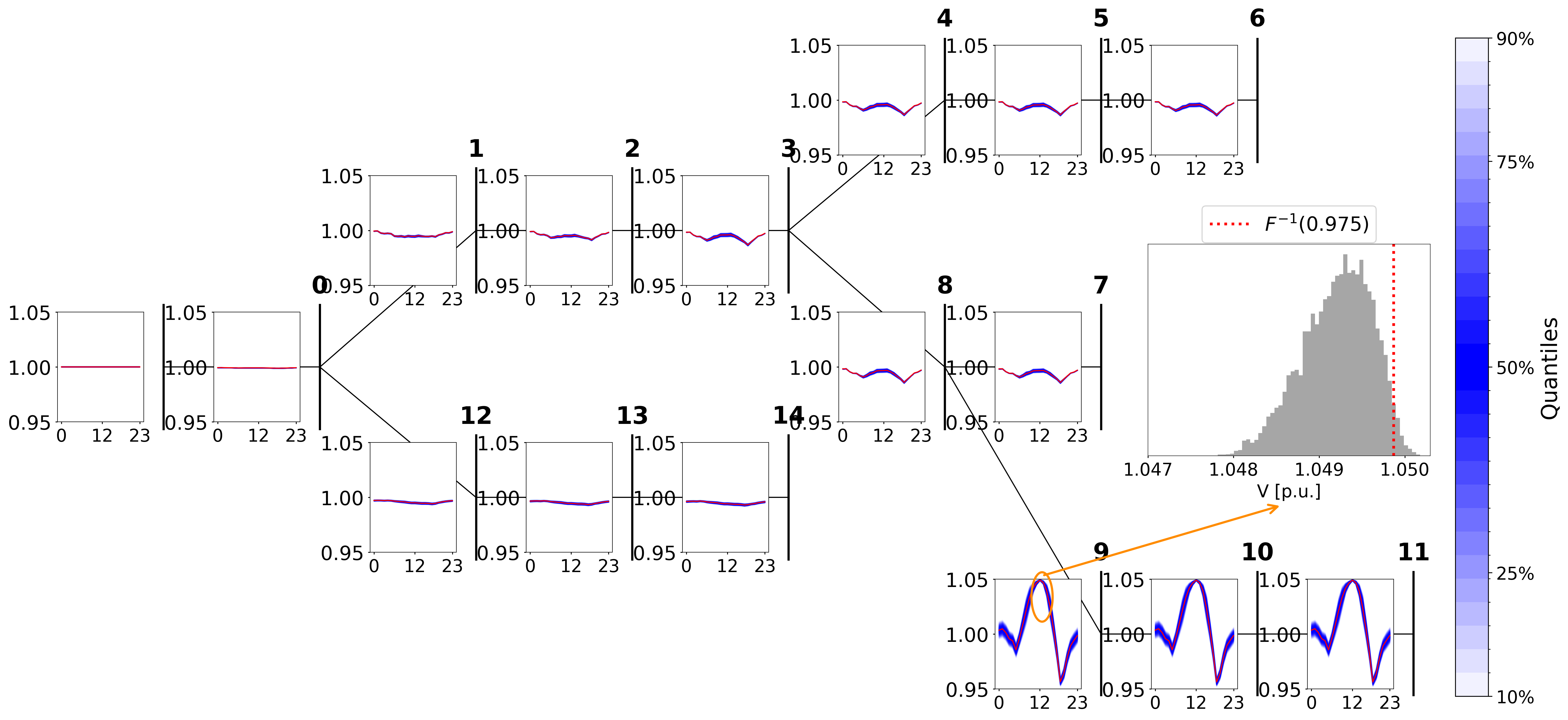}
    \caption{Distribution of nodal voltage magnitudes and the topology of the considered grid. The probabilistic evolution of the nodal voltage magnitudes is shown above every bus. The x-axis represents the 24 timesteps in a day and the y-axis represents the nodal voltage magnitude. All axes have the same scaling. The inserted histogram shows the distribution of the nodal voltage in bus 9 for the 12-th timestep, when the network experiences voltage congestion due to increased \ac{PV} production.}
    \label{fig:case3-voltages}
\end{figure*}

The probabilistic constraint expressed using the variance of the nodal voltage magnitude allows the market clearing model to account for uncertainty in the day-ahead stage. If the distribution is Gaussian, the chance-constrained formulation allows to express constraints in terms of quantiles, i.e. the probability of voltage violation can be accurately modeled. If distributions are very different from Gaussian, a distributionally robust constraint can be considered \cite{kuhn_distributionally_2025}. \cref{fig:case3-voltages} shows the distribution of voltage magnitudes in all buses. The voltage magnitude in the slack bus is considered to be fixed at 1 p.u. The optimal setpoint results in binding \textit{probabilistic} voltage constraints in bus 9 during midday, when the \ac{PV} production is at a maximum. As can be observed, some of the realizations of the voltage magnitude are higher than 1.05 p.u., however, the probabilistic constraint guarantees a certain level of confidence, i.e. only a certain percentage of scenarios result in constraint violation. 

\begin{figure*}[t]
    \centering
    \includegraphics[width=0.95\textwidth]{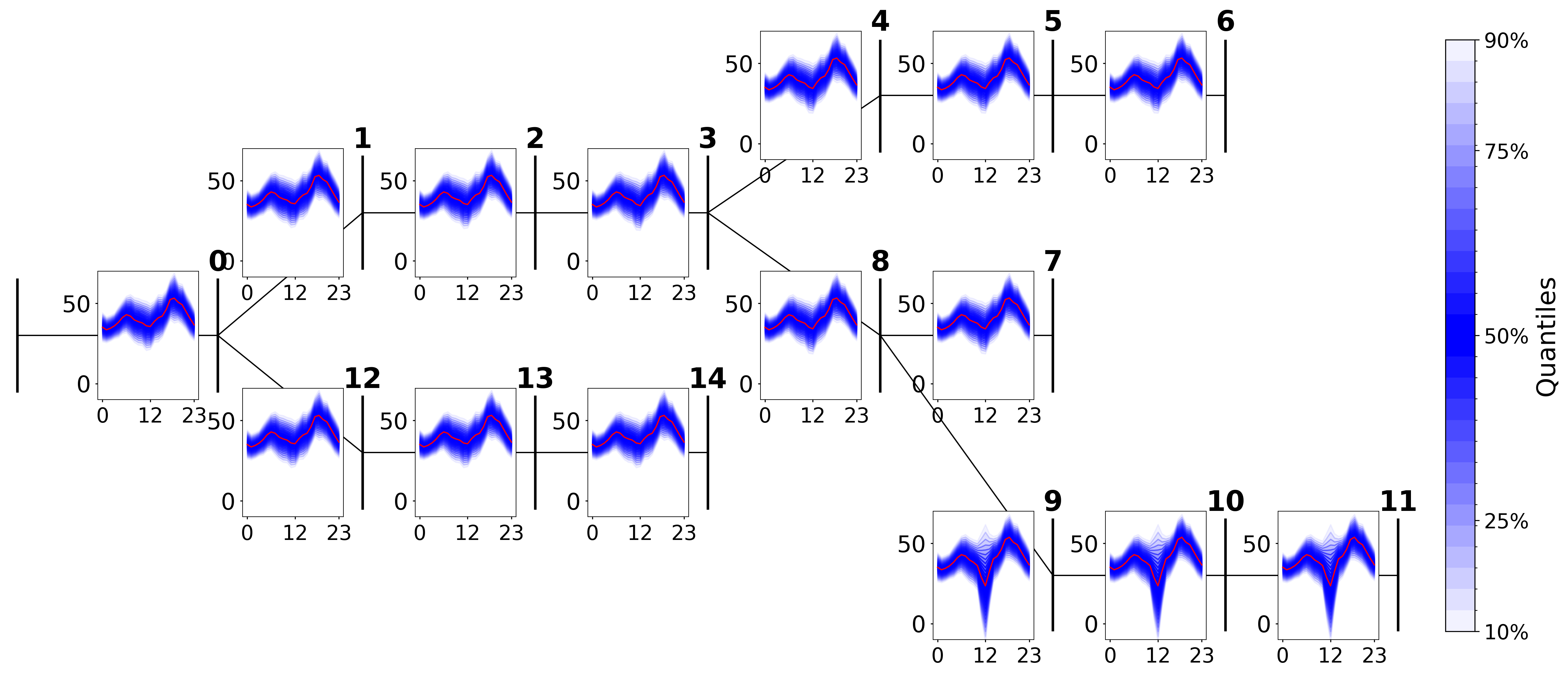}
    \caption{Distribution of \acp{PLMP} and the topology of the considered grid. The probabilistic evolution of the \acp{PLMP} is shown above every bus. The x-axis represents the 24 timesteps in a day and the y-axis represents the \acp{PLMP}. All axes have the same scaling.}
    \label{fig:case3-lmps}
\end{figure*}

\subsubsection{Impact of passive balancing}
\begin{figure}[H]
    \centering
    \includegraphics[width=0.50\textwidth]{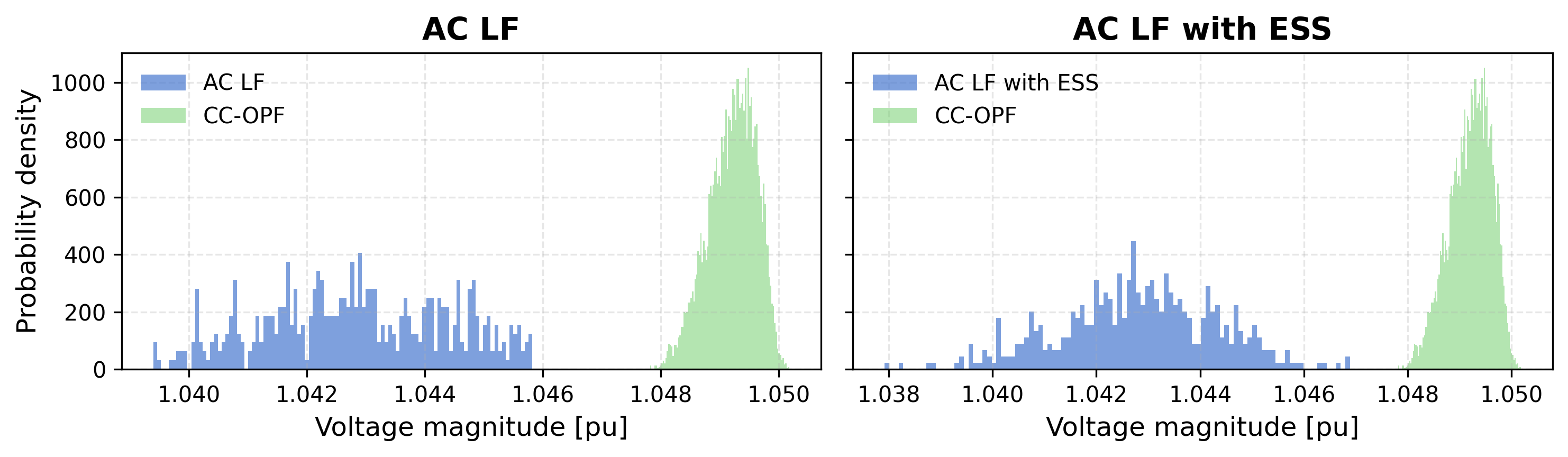}
    \caption{Distribution of nodal voltage magnitude in the binding bus at timestep 12.}
    \label{fig:case3-voltage_node10}
\end{figure}

We proceed to simulate two flexible passive end-prosumers located in buses 5 and 10. The end-prosumer located in bus 10 are experiencing the high volatility of realtime prices caused by the voltage congestion. Considering a strategic end-prosumer aiming to maximize their profit, we employ the \ac{DP}-based optimizer to simulate their actions and impact on the grid. As can be seen in \cref{fig:case3-voltage_node10}, the actions of the flexible prosumer do not negatively impact the bus voltage, calculated through an a posteriori AC load flow. We also observe that, the lindistflow equations approximate the nodal voltages, by linearizing the nonlinear power flow equations and neglecting shunts. This linearization leads to dropping of the terms related to losses \cite{baran_network_1989}.

\subsection{Case IV: 179-bus system}

\begin{figure}[H]
    \centering
    \includegraphics[width=0.5\textwidth]{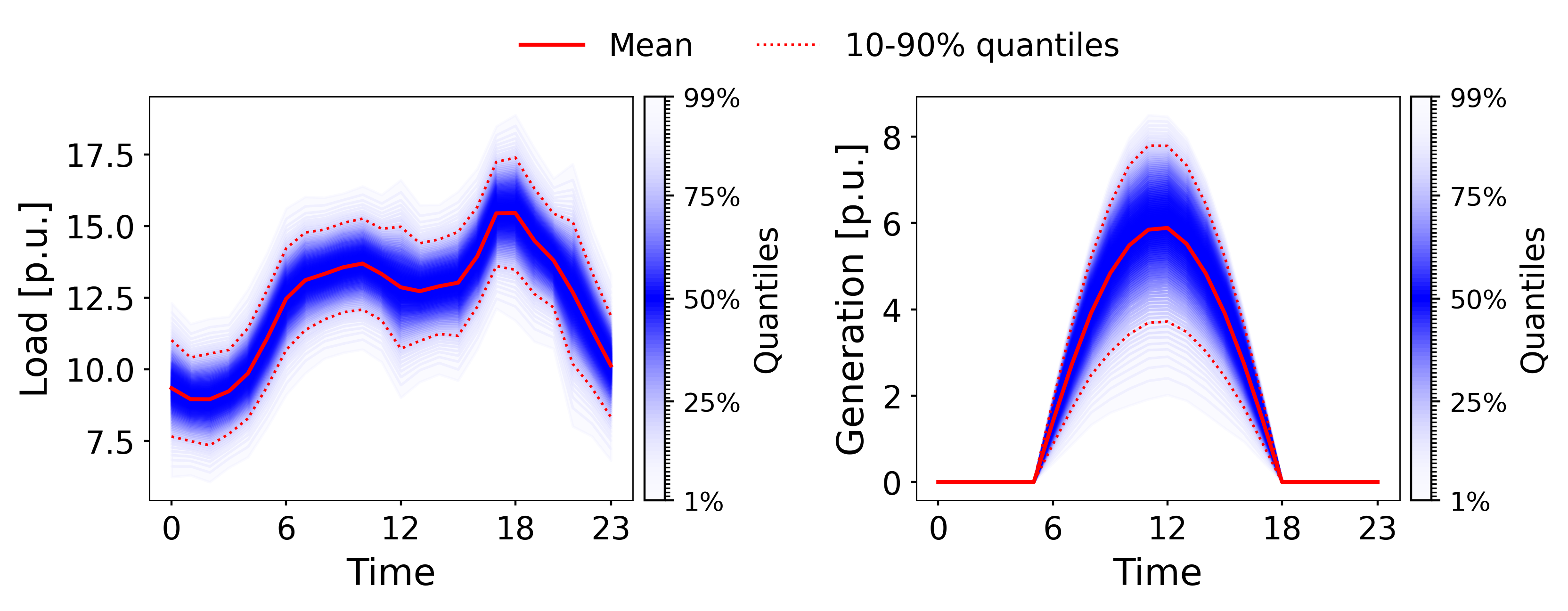}
    \caption{Aggregate load and \ac{PV} profiles for case study IV. The loads are modeled using a mix of gaussian and beta-distributions, while the \ac{PV} is modeled using beta-distributions.}
    \label{fig:case135-config}
\end{figure}

The final case-study considers a larger grid based on the 179-bus Oberrhein \ac{MV} network \cite{pandapower_mv_nodate}. We consider a configuration with 146 distributed loads and 153 \ac{PV} plants as well as two flexible generators in buses 0 (slack) and 70. Loads are modeled as a mix of gaussian distributions and beta distributions with shape parameters $\alpha=5,\beta=2$ and $\alpha=4,\beta=2$, while the \ac{PV} generation is modeled as beta distributions with shape parameters $\alpha=5$ and $\beta=2$. An example profile for both load and \ac{PV} generation is shown in \cref{fig:case135-config}. The purpose of this case study is to highlight the scalability of the proposed methodology. \cref{tab:computational-time} summarizes the computational time on the 14-bus and 179-bus networks and shows that it remains modest, even as grids grow larger. The principal factors impacting computational time is network size and number of uncertain drivers. The number of uncertain drivers can be kept low, by considering common influential factors, for example temperature for load demand or solar irradiation for \ac{PV} production \cite{muhlpfordt_chance-constrained_2019}.

\begin{figure*}[t]
    \centering

    \begin{subfigure}[t]{0.99\textwidth}
        \centering
        \includegraphics[width=\textwidth]{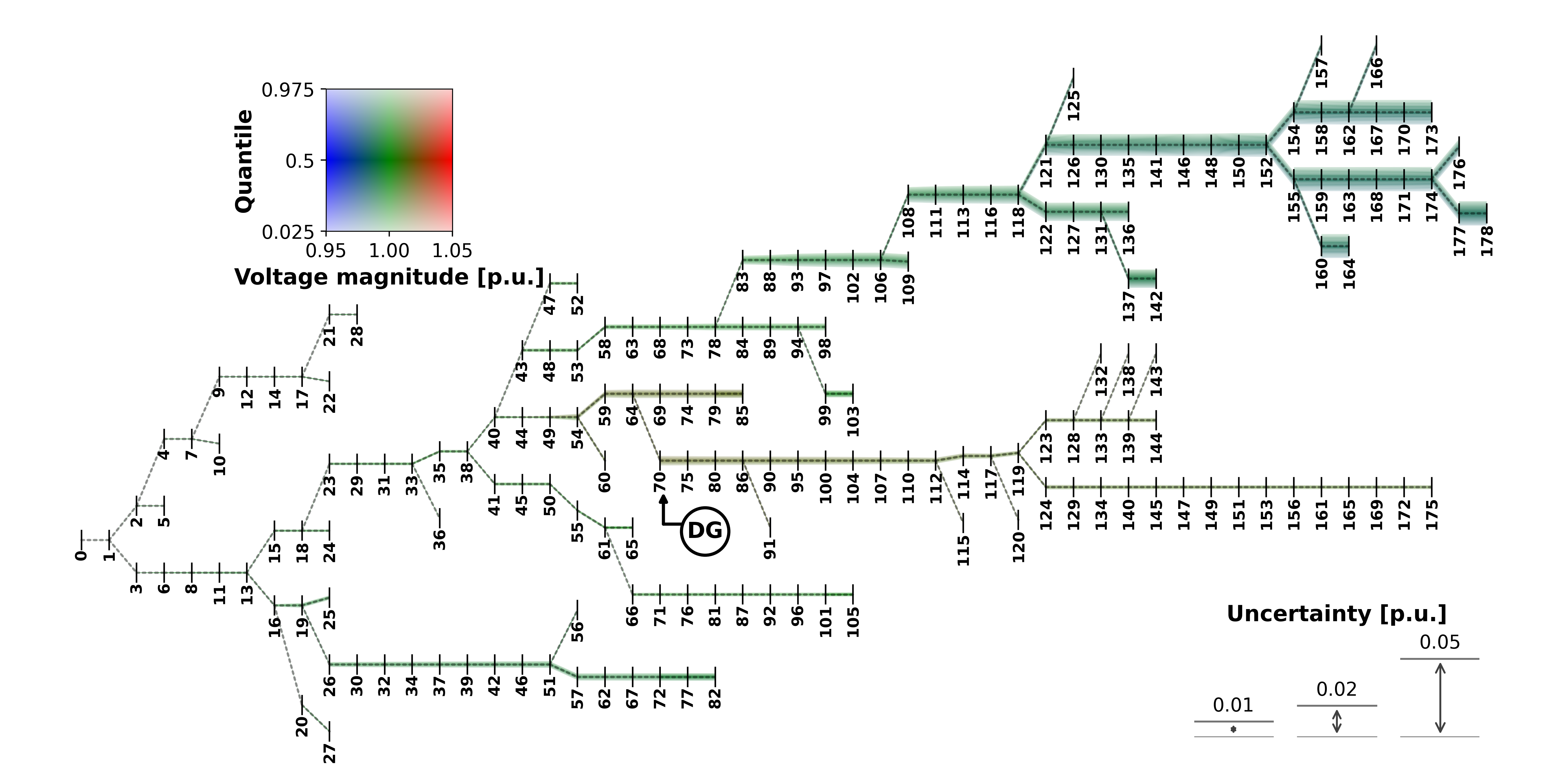}
        \caption{Distribution of bus voltage magnitudes.}
        \label{fig:case135-v0}
    \end{subfigure} \\
    \hfill
    \begin{subfigure}[t]{0.99\textwidth}
        \centering
        \includegraphics[width=\textwidth]{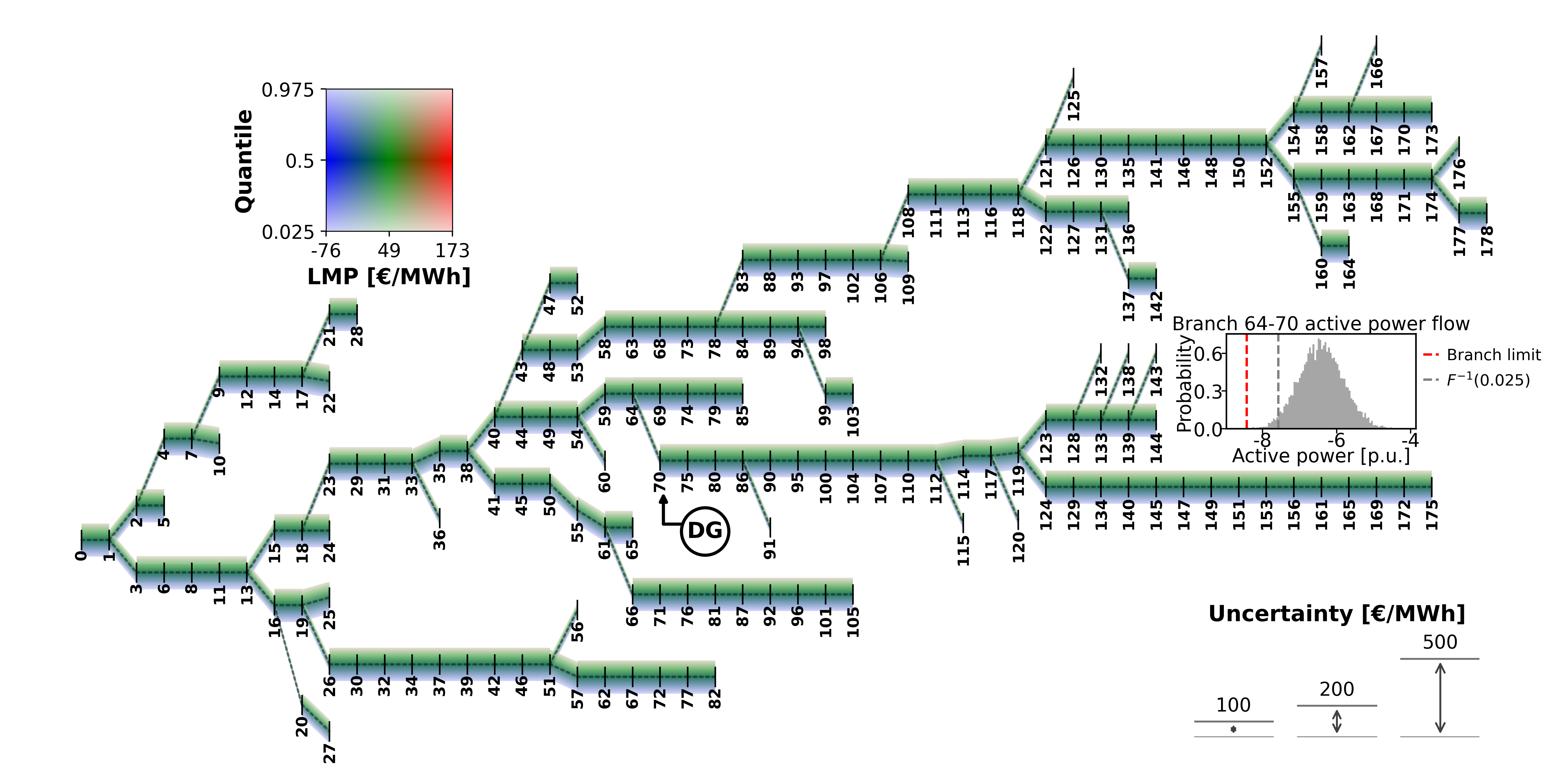}
        \caption{Distribution of \acp{LMP}.}
        \label{fig:case135-prices0}
    \end{subfigure}

    \caption{Distribution of nodal voltage magnitudes and \acp{PLMP} across the network for hour~0. The distributed generator is connected to bus~70. The dotted black lines represent the mean value per bus.}
    \label{fig:case135-ts0}
\end{figure*}

\begin{figure*}[t]
    \centering

    \begin{subfigure}[t]{0.99\textwidth}
        \centering
        \includegraphics[width=\textwidth]{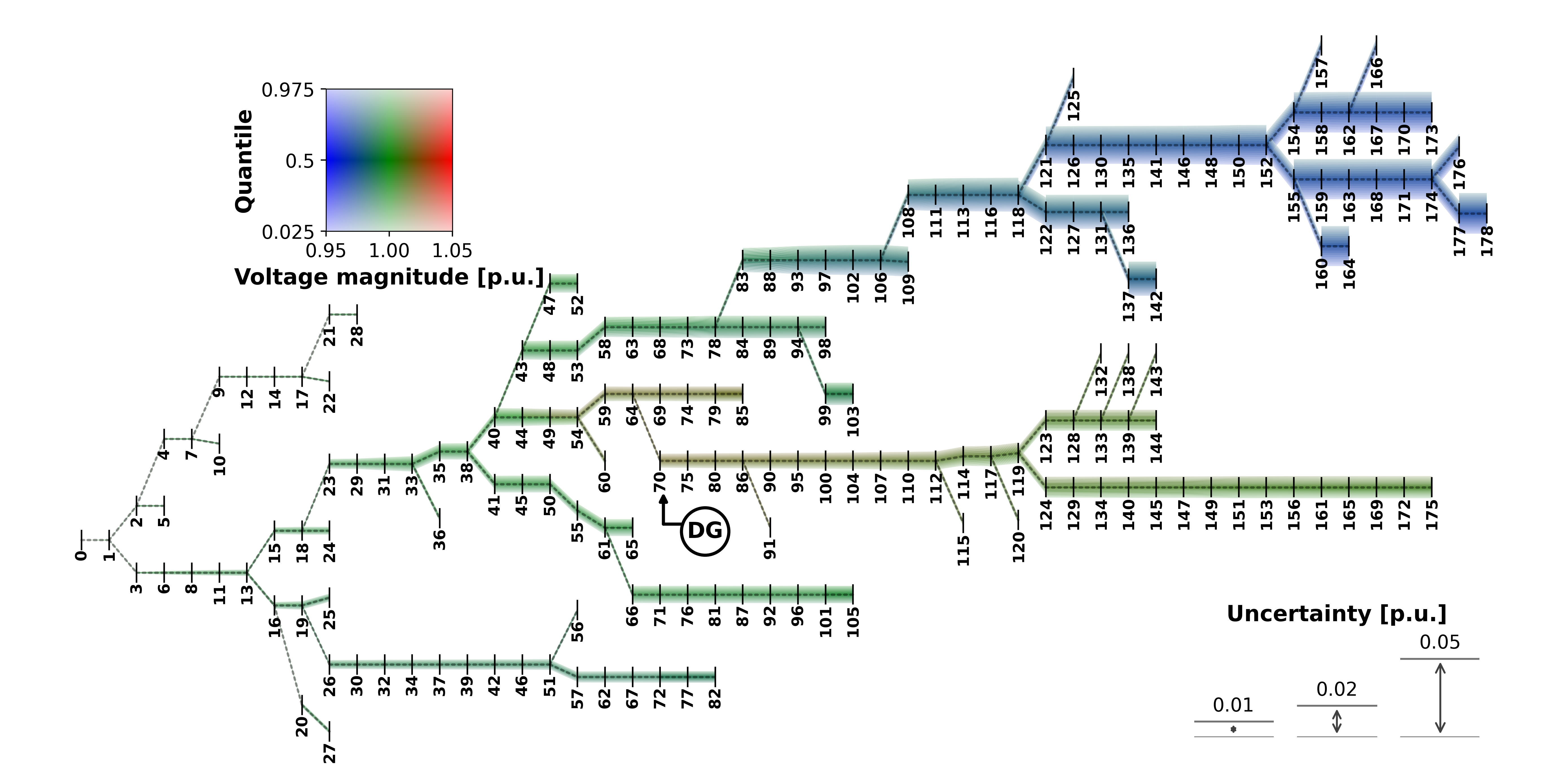}
        \caption{Distribution of bus voltage magnitudes.}
        \label{fig:case135-v18}
    \end{subfigure} \\
    \hfill
    \begin{subfigure}[t]{0.99\textwidth}
        \centering
        \includegraphics[width=\textwidth]{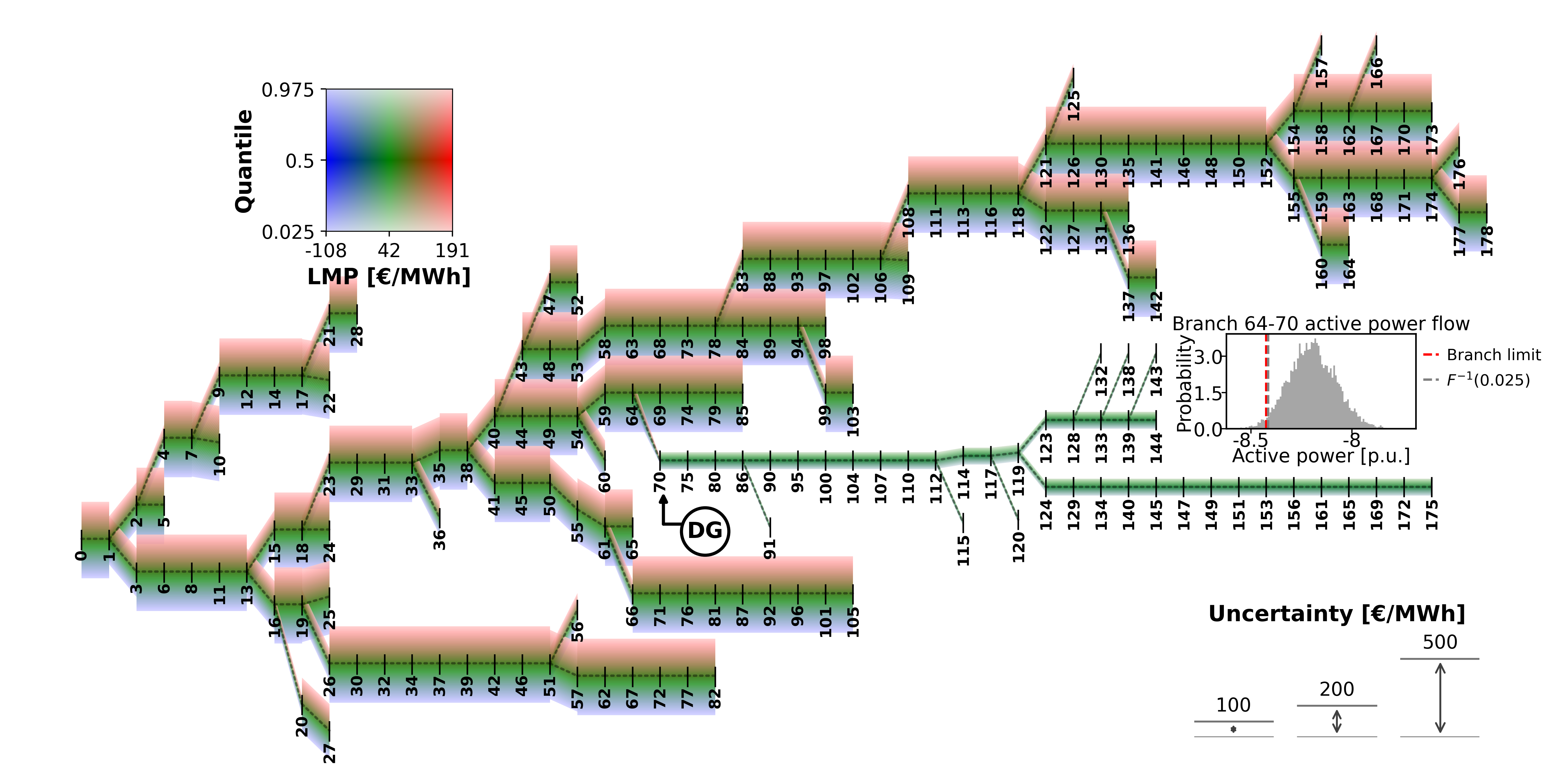}
        \caption{Distribution of \acp{PLMP}.}
        \label{fig:case135-prices18}
    \end{subfigure}

    \caption{Distribution of nodal voltage magnitudes and \acp{PLMP} across the network for hour~19. The distributed generator is connected to bus~70. The dotted black lines represent the mean value per bus.}
    \label{fig:case135-ts18}
\end{figure*}

Results for the 179-bus system are shown in \cref{fig:case135-ts0,fig:case135-ts18} for hour 0 and hour 19, respectively. These time steps refer to midnight load and afternoon peak load. \cref{fig:case135-v0,fig:case135-v18} shows the distribution of bus voltage magnitudes for the two time steps. As can be observed, the uncertainty is lower at night and higher during the afternoon peak. This is directly reflected in \acp{PLMP} as shown in \cref{fig:case135-prices0,fig:case135-prices18} where the higher load and uncertainty in the afternoon peak translates to higher spread of \acp{PLMP}. During the evening peak, branch 64-70 reaches its capacity creating a congestion in the system. The flexible generator connected to bus 70 becomes the marginal generator in the congested zone, reflecting a difference of \acp{PLMP} between the congested and uncongested parts of the network.

\begin{table}[!htbp]
  \centering
  \caption{Computational time for the 24-period \ac{gPC} \ac{CC-OPF} using the lindistflow equations for the 14-bus and 179-bus networks. Solved on a standard laptop with a 12th generation Intel Core i9 CPU and 32 GB of RAM. Only the time reported by Gurobi is counted.}
  \label{tab:computational-time}
  \renewcommand{\arraystretch}{1.2}   % a bit more breathing-room
  \begin{tabular}{ccl}
    \toprule
    & 14-bus & 179-bus\\
    \midrule
    Computational time [s] & 3.76 & 403\\
    \bottomrule
  \end{tabular}
\end{table}

\section{Discussion and conclusion}\label{sec:discussion}
%\acresetall
This study proposes an optimization-based \ac{LEM} that naturally integrates non-gaussian probability distributions and expresses prices probabilistically. The proposed method overcomes existing limitations of \acp{LEM} by allowing end-prosumers participation through passive balancing, i.e. not requiring active bidding or communication from end-consumers to the \ac{MO}. From the \acp{MO} point of view, the proposed methodology ensures grid constraints are satisfied probabilistically and actions from rational market participants help reduce risk of congestion. Only measurements of the considered uncertain drivers are necessary during realtime operation, effectively limiting the need for situational awareness from the market/grid operator. During the day-ahead scheduling phase, prices are represented as probability distributions, effectively allowing market participants to develop advanced optimization algorithms to optimally dispatch their resources following the risk of large uncertainty in realtime prices. During realtime, the \ac{MO} measures the state of the grid and the modeled meteorological variables, i.e. the realization of the uncertainty. With these measurements, the actual realtime price is calculated by simply evaluating the polynomial chaos expansion and without the need to rerun the optimization model. Prices are then communicated to end-prosumers and valid for the market time unit. Power grid constraints are accounted for by using the lindistflow equations and constraints are reformulated as chance-constraints, allowing for a tunable probability of violation. 

The computational aspects are of particular importance for local markets as there are many, small market participants and limited operational capacity for DSOs/aggregators. By solving both the day-ahead and the realtime optimal market problem in the day-ahead stage, we avoid strict time-constraints and allow for remedial actions, if needed, by the \ac{MO} well in advance of realtime. \cref{tab:computational-time} show that the time required to solve the daily optimization problem remains modest, even for larger grids. Compared to peer-to-peer markets, our approach does not require active participation by end-prosumers during market clearing. Instead, they are incentivized to perform passive balancing, i.e., shift or adapt their consumption as a reaction to real-time prices. An obvious limitation of our approach is that excessive passive balancing by end-prosumers can lead to oscillations in the balancing state of the system, reducing system stability. This however, requires significant volumes of flexible assets. Practical measures to reduce this risk, such as delayed publication of system data or ramp rates, have been proposed and studied \cite{lips_insights_2025}.

Another important aspect is accounting for realistic physical constraints in distribution grids. Compared to central electricity markets, which rely on simple assumptions such as zonal pricing or nodal pricing using the DC-approximation, distribution grids must better account for binding voltage constraints. The lindistflow grid model is therefore a more suitable choice for local electricity markets. At the same time, the proposed framework does not require the \ac{MO} to know how much and where energy storage assets are installed as their control is performed by the end-prosumers.

Ultimately, achieving large-scale participation of end-prosumers in flexibility provision necessitates simplicity, as this is essential to fostering both the adoption and the effectiveness of \acp{LEM}. This framework allows for simple automatic trading by end-prosumers, for example through a HEMS, which can read the price signals from the \ac{MO} and optimize the control of flexible units. In this respect, we argue that exposing end-prosumers to realtime variable electricity prices is an incentive to increase their participation in flexibility services. This in turn leads to a more efficient operation of the electricity grid and a reduction of volatile and extreme prices.

\section*{Funding}
This research is carried out in the frame of the project “UrbanTwin: An urban digital twin for climate action: Assessing policies and solutions for energy, water and infrastructure” with the financial support of the ETH-Domain Joint Initiative program in the Strategic Area Energy, Climate and Sustainable Environment.

\appendices
\section*{Case study parameters}
The overview of the parameters used for the case studies is found in \cref{tab:case_params}.
\begin{table*}[!t]
  \centering
  \caption{Case study parameters. Numbers in parenthesis under stochastic prosumption refers to the index of the stochastic germ. A detailed overview of case study parameters and locations of loads and PV-plants for the 179-bus system can be found in the Supplementary Data.}
  \label{tab:case_params}
  \renewcommand{\arraystretch}{1.2}   % a bit more breathing-room
  \begin{tabular}{ccccl}
    \toprule
    \multicolumn{5}{c}{\textbf{Stochastic germ}}\\
    \midrule
    Index & \multicolumn{2}{c}{Distribution} & Polynomial basis & \\
    \midrule
    1 & \multicolumn{2}{c}{$\mathcal{N}(0,1)$} & Hermite  \\
    2 & \multicolumn{2}{c}{$\mathrm{B}\!\left([0,1],\,5,\,2\right)$} & Jacobi \\
    3 & \multicolumn{2}{c}{$\mathrm{B}\!\left([0,1],\,4,\,2\right)$} & Jacobi \\
    %---------------------------------------------------------
    \midrule
    \multicolumn{5}{c}{\textbf{Case Study I}}\\
    \midrule
Bus & \multicolumn{2}{c}{Stochastic prosumption} & Flexible generation & Cost parameters\\
    \cmidrule(lr){2-3}            
      & Load & Generation (PV) & \\
    \midrule
    0 (slack) &  &  &  X  & $c=50$, $C=15$, $C_2=200$\\
    3,4,7,9 & X (1) &  &  &  \\
    1,5,6,8,10,11,12,13,14 & X (2) &  &  &    \\
    3,4,5,6,8,9,10,11 &  & X (3) &  &    \\
    %---------------------------------------------------------
    \midrule
    \multicolumn{5}{c}{\textbf{Case Study II}}\\
    \midrule
Bus & \multicolumn{2}{c}{Stochastic prosumption} & Flexible generation & Cost parameters\\
    \cmidrule(lr){2-3}            
      & Load & Generation (PV) & \\
    \midrule
    0 (slack) &  &  &  X & $c=20$, $C=15$, $C_2=100$ \\
    3,4,7,9 & X (1) &  &       \\
    1,5,6,8,10,11,12,13,14 & X (2) &  &       \\
    3,4,5,6,8,9,10,11 &  & X (3) &       \\
    9 &  &  &  X  & $c=100$, $C=15$, $C_2=20$   \\
    %---------------------------------------------------------
    \midrule
    \multicolumn{5}{c}{\textbf{Case Study III}}\\
    \midrule
Bus & \multicolumn{2}{c}{Stochastic prosumption} & Flexible generation & Cost parameters\\
    \cmidrule(lr){2-3}            
      & Load & Generation (PV) & \\
    \midrule
    0 (slack) &  &  &  X  & $c=10$, $C=5$, $C_2=10$ \\
    9 &  &  &  X    & $c=0$, $C=1000$, $C_2=500$ \\
    1,3,4,5,6,7,8,9,10,11,12,13,14 & X (1) &  &       \\
    3,4,5,6,8,9,10,11 &  & X (3) &       \\
    \midrule
    \multicolumn{5}{c}{\textbf{Case Study IV}}\\
    \midrule
Bus & \multicolumn{2}{c}{Stochastic prosumption} & Flexible generation & Cost parameters\\
    \cmidrule(lr){2-3}            
      & Load & Generation (PV) & \\
    \midrule
    0 (slack) &  &  &  X  & $c=1$, $C=20$, $C_2=100$\\
    70 &  &  &  X & $c=10$, $C=5$, $C_2=50$ \\
    Various &  & X (3) &       \\
    Various & X (1) &  &       \\
    Various & X (2) &  &       \\
    \bottomrule
  \end{tabular}
\end{table*}

% \section{References Section}
% You can use a bibliography generated by BibTeX as a .bbl file.
%  BibTeX documentation can be easily obtained at:
%  http://mirror.ctan.org/biblio/bibtex/contrib/doc/
%  The IEEEtran BibTeX style support page is:
%  http://www.michaelshell.org/tex/ieeetran/bibtex/
 
%  % argument is your BibTeX string definitions and bibliography database(s)
% %\bibliography{IEEEabrv,../bib/paper}
% You can manually copy in the resultant .bbl file and set second argument of $\backslash${\tt{begin}} to the number of references
%  (used to reserve space for the reference number labels box).

\bibliographystyle{IEEEtran} 
\bibliography{references}

\vfill

\end{document}